\documentclass[12pt]{article}
\usepackage[dvips]{graphics,color}

\usepackage{epsfig}

\usepackage{subfigure} % Written by Steven Douglas Cochran
\usepackage{url}       % Written by Donald Arseneau
\usepackage{amsmath}   % From the American Mathematical Society
\usepackage{amssymb}%for sets R,Z,C. E.g. \mathbb gives wide Blackboard font

\usepackage{setspace}

\newtheorem{definition}{Definition}%[within]

\newtheorem{proof}{Proof}

\newtheorem{theorem}{Theorem}
\newcommand{\xv}{{\bf x}}   %bold face x, vector
\newcommand{\yv}{{\bf y}}   %bold face x, vector
\newcommand{\Hm}{{\bf H}}   %bold face x, vector
\newcommand{\E}{{\mathcal E}}

\newcommand{\nchoosek}[2]{\left(\begin{array}{c}#1\\#2\end{array}\right)}

\begin{document}
%
% paper title
\title{A General Method for Finding Low Error Rates of LDPC Codes}
%
%
% author names and IEEE memberships
% note positions of commas and nonbreaking spaces ( ~ ) LaTeX will not break
% a structure at a ~ so this keeps an author's name from being broken across
% two lines.
% use \thanks{} to gain access to the first footnote area
% a separate \thanks must be used for each paragraph as LaTeX2e's \thanks
% was not built to handle multiple paragraphs
\author{\begin{tabular}[t]{c@{\extracolsep{8em}}c}
Chad A. Cole  &  Eric. K. Hall \\
Stephen G. Wilson & Thomas R. Giallorenzi \\
\\
%I. M. Anonymous  & M. Y. Coauthor \\
%Dept. of Elec and Comp. Engr. & L-3 Communications \\
Univ. of Virginia &  L-3 Communications \\
Charlottesville, VA 22904 & Salt Lake City, UT 84116
\thanks{This work is supported by L-3 Communications.
This work has been submitted to the IEEE for possible publication.
Copyright may be transferred without notice, after which this
version may no longer be accessible.}
\end{tabular}}
\markboth{Journal of \LaTeX\ Class Files,~Vol.~1, No.~11,~November~2002}{Shell \MakeLowercase{\textit{et al.}}: Bare Demo of IEEEtran.cls for Journals}
% The only time the second header will appear is for the odd numbered pages
% after the title page when using the twoside option.
%
% *** Note that you probably will NOT want to include the author's name in ***
% *** the headers of peer review papers.                                   ***

% If you want to put a publisher's ID mark on the page
% (can leave text blank if you just want to see how the
% text height on the first page will be reduced by IEEE)
%\pubid{0000--0000/00\$00.00~\copyright~2002 IEEE}

% use only for invited papers
%\specialpapernotice{(Invited Paper)}

% make the title area
\maketitle
\newpage

\begin{abstract}
This paper outlines a three-step procedure for determining the low
bit error rate performance curve of a wide class of LDPC codes of
moderate length.  The traditional method to estimate code
performance in the higher SNR region is to use a sum of the
contributions of the most dominant error events to the probability
of error.  These dominant error events will be both code and decoder
dependent, consisting of low-weight codewords as well as
non-codeword events if ML decoding is not used.  For even moderate
length codes, it is not feasible to find all of these dominant error
events with a brute force search. The proposed method provides a
convenient way to evaluate very low bit error rate performance of an
LDPC code without requiring knowledge of the complete error event
weight spectrum or resorting to a Monte Carlo simulation. This new
method can be applied to various types of decoding such as the full
belief propagation version of the message passing algorithm or the
commonly used min-sum approximation to belief propagation. The
proposed method allows one to efficiently see error performance at
bit error rates that were previously out of reach of Monte Carlo
methods. This result will provide a solid foundation for the
analysis and design of LDPC codes and decoders that are required to
provide a guaranteed very low bit error rate performance at certain
SNRs.

Keywords - LDPC codes, error floors, importance sampling.

\end{abstract}

%This work has been submitted to the IEEE for possible publication.
%Copyright may be transferred without notice, after which this
%version may no longer be accessible.

%\begin{keywords}
%LDPC codes, error floors, importance sampling.
%\end{keywords}
% Note that keywords are not normally used for peerreview papers.

% For peer review papers, you can put extra information on the cover
% page as needed:
% \begin{center} \bfseries EDICS Category: 3-BBND \end{center}
%
% For peerreview papers, inserts a page break and creates the second title.
% Will be ignored for other modes.
%\IEEEpeerreviewmaketitle

\section{Introduction}
% The very first letter is a 2 line initial drop letter followed
% by the rest of the first word in caps.
%
% form to use if the first word consists of a single letter:
% \PARstart{A}{demo} file is ....
%
% form to use if you need the single drop letter followed by
% normal text (unknown if ever used by IEEE):
% \PARstart{A}{}demo file is ....
%
% Some journals put the first two words in caps:
% \PARstart{T}{his demo} file is ....
%
% Here we have the typical use of a "T" for an initial drop letter
% and "HIS" in caps to complete the first word.
%\PARstart{T}{his} demo file is intended to serve as a ``starter file"
%for IEEE journal papers produced under \LaTeX\ using IEEEtran.cls version
%1.6b and later.
% You must have at least 2 lines in the paragraph with the drop letter
% (should never be an issue)
% May all your publication endeavors be successful.

The recent rediscovery of the powerful class of codes known as
low-density parity-check (LDPC) codes \cite{Gallager, mackay-99} has
sparked a flurry of interest in their performance characteristics.
Certain applications for LDPC codes require a guaranteed very low
bit error rate, and there is currently no practical method to
evaluate the performance curve in this region. The most difficult
task of determining the error `floor' of a code (and decoder) in the
presence of additive white Gaussian noise (AWGN) is locating the
dominant error events that contribute most of the error probability
at high SNR. Recently, a technique for solving this problem for the
class of moderate length LDPC codes \cite{cole-06} has been
discovered. Since an ML detector is not commonly used (or even
feasible) for decoding LDPC codes, most of these error events are a
type of non-codeword error called \emph{trapping sets} (TS)
\cite{richardson}.  Since the error contribution of a TS is not
given by a simple Q-function, as in the case of the two-codeword
problem for an ML decoder, it is not clear how the list of error
events (mainly TS) returned by the search technique of
\cite{cole-06}, henceforth referred to as the `decoder search,'
should be utilized to provide a complete picture of a code's low bit
error rate performance.
    This paper will present a three-pronged attack for determining
the complete error performance of a short to moderate length LDPC
code for a variety of decoders.  The first step is to utilize the
decoder search to build a list of the dominant error events. Next, a
deterministic noise is directed along a line in $n$-dimensional
space towards each of these dominant events and a Euclidean distance
to the error boundary is found by locating the point at which the
decoder fails to converge to the correct state. This step will be
crucial in determining which of the error events in our initial list
is truly dominating (i.e. nearest in $n$-dimensional decoding space
to our reference all-zeros codeword). The final step involves an
importance sampling (IS) \cite{Srinivasan-02, Smith-97, wu-95}
procedure for determining the low bit error rate performance of the
entire code. The IS technique has been applied to LDPC codes before
\cite{Xia-03, cavus-05, Holzlohner-05} with limited success. The
method proposed in \cite{Xia-03} does not scale well with block
length.  The method in \cite{cavus-05} uses IS to find the error
contribution of each TS individually, and then a sum of these error
contributions gives the total code performance.  This method is
theoretically correct, but it has a tendency to underestimate the
error curves since inevitably some important error events will not
be known. The method proposed in this paper is effective for block
length $n < 10000$ or so, and does not require that the initial list
of dominant TS be complete, thus improving upon some of the
limitations of previous methods of determining very low bit error
rates.

This paper is organized as follows: Section \ref{sec: terminology}
introduces some terms and concepts necessary to understand the
message passing decoding of LDPC codes and what causes their error
floors. Section \ref{sec: decoder_search} gives a self-contained
introduction to the decoder search procedure of \cite{cole-06}.
Section \ref{sec: error_boundary} details step two of our general
procedure - the localized decoder search for the error boundary of a
given TS. Section \ref{sec: IS} reviews the basics of importance
sampling (step three of our procedure). Section \ref{sec: procedure}
puts together all three steps of our low bit error performance
analysis method and gives a step-by-step example. Section \ref{sec:
results} gives some results for different codes and decoders and
shows the significant performance differences in the low bit error
region of different types of LDPC codes. The final section
summarizes the contribution offered in this paper.

\section{Preliminaries}\label{sec: terminology}

LDPC codes, the revolutionary form of block coding that allows large
block length codes to be practically decoded at SNR's close to the
channel capacity limit, were first presented by Gallager in the
early 1960's \cite{Gallager}. These codes have sparse parity check
matrices, denoted by ${\bf H}$, and can be conveniently represented
by a \textit{Tanner graph} \cite{tanner-81}, where each row of ${\bf
H}$ is associated with a \textit{check node} $c_i$, each column of
${\bf H}$ is associated with a \textit{variable node} $v_j$, and
each position where $H_{ij} = 1$ defines an edge connecting $v_i$ to
$c_i$ in the graph.  A \textit{regular} $\{j,k\}$ graph has $j$ `1's
per column and $k$ `1's per row.

The iterative message passing algorithm (MPA), also referred to as
Belief Propagation when using full-precision soft data in the
messages, is the method commonly used to decode LDPC codes.  This
algorithm passes messages between variable and check nodes,
representing the probability that the variable nodes are `1' or `0'
and whether the check nodes are satisfied.  The following equations,
representing the calculations at the two types of nodes, will be
considered in the log likelihood ratio (LLR) domain with notation
taken from \cite{ryan}.

We will consider BPSK modulation on an AWGN memoryless channel,
where each received channel output $y_i = \sqrt{E_s}x_i + n_i$ is
conditionally independent of any others. The transmitted bits,
$x_i$, are modulated by $0 \rightarrow +1, 1 \rightarrow -1$ and are
assumed to be equally likely.  $n_i \backsim N(0, N_o/2)$ is the
noise with two-sided PSD $N_o/2$.  The a posteriori probability for
bit $x_{i}$, given the channel data, $y_{i}$, is given by

\begin{align} \label{eq:LDPC.channeldata}
P(x_{i} | y_{i}) = \frac{P(y_i | x_i)P(x_i)}{P(y_i)}
\end{align}

The LLR of the channel data for the AWGN case is denoted $Lc_i =
\log\frac{Pr(x_i=0| y_i)}{Pr(x_i=1| y_i)} = 4y_iE_s/N_o$.

The LLR message from the $j^{th}$ check node to the $i^{th}$
variable node is given by

\begin{align} \label{eq:LDPC.checkmsg}
Lr_{ji} = 2\tanh^{-1}\left[\prod\limits_{V_j \backslash i}
\tanh(Lq_{ij}/2)\right]
\end{align}

The set $V_j$ is all of the variable nodes connected to the $j^{th}$
check node and $C_i$ is all of the check nodes connected to the
$i^{th}$ variable node.  $V_j \backslash i$ is the set $V_j$ without
the $i^{th}$ member, and $C_i \backslash j$ is likewise defined. The
LLR message from the $i^{th}$ variable node to the $j^{th}$ check
node is given by

\begin{align} \label{eq:LDPC.varmsg}
  Lq_{ij} = \sum\limits_{C_i \backslash j} Lr_{ji} + Lc_{i}
\end{align}

The marginal LLR for the $i^{th}$ code bit, which is used to make a
hard decision for the transmitted codeword is

\begin{align} \label{eq:LDPC.marginalmsg}
 LQ_{i} = \sum\limits_{C_i} Lr_{ji} + Lc_{i}
\end{align}

If $LQ_{i} > 0$, then $\hat{x}_i = 0$, else $\hat{x}_i = 1$. The
decoder continues to pass these messages in each iteration until a
preset maximum number of iterations is reached or the estimate ${\bf
\hat{x}} \in \mathcal{C}$, the set of all valid codewords. For a
more detailed exposition of the MPA, see \cite{ryan}.

\begin{figure} [h]
\begin{center}
\epsfig{file=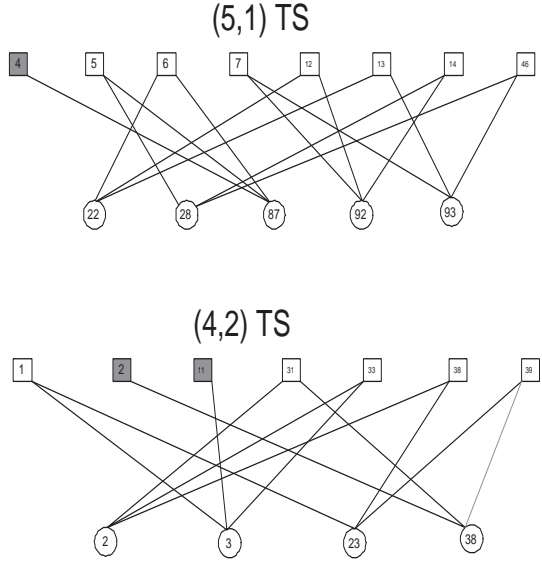, height=3.0in, width=2.8in} \caption{A (5,1) and
(4,2) TS subgraph \label{fig: Error.TS_5_1}}
\end{center}
\end{figure}

As with all linear codes, it is convenient to assume the all-zeros
codeword as a reference vector for study.  If a subset of variable
nodes and check nodes is considered, members of this subset will be
called the \textit{active} nodes of a \textit{subgraph}.  By
studying the local subgraph structure around each variable node,
arguments can be made about the global graph structure.  For
example, bounds can be placed on the minimum distance of a code by
only considering the local structure of a code \cite{tanner-81}.
 Small subsets of non-zero bits which do not form a valid codeword
but still cause the MPA decoder problems are well-documented in the
literature and typically referred to as \textit{trapping sets} (TS)
\cite{richardson}. A TS, $\bf{x}$, is a length-$n$ bit vector
denoted by a pair $(a,b)$, where $a$ is the Hamming weight of the
bit vector and $b$ is the number of unsatisfied checks, i.e. the
Hamming weight of the syndrome $\bf{xH}^{T}$.  Alternatively, from a
Tanner graph perspective, a TS could be defined as the $a$ nonzero
variable nodes of $\bf{x}$ and all of the check nodes connected by
one edge to those $a$ variable nodes.  A valid codeword is a TS with
$b=0$. Two examples of TS, both extracted from a (96,48) code on
MacKay's website \cite{Mackay_codes}, are shown in Figure \ref{fig:
Error.TS_5_1}. The shaded check nodes, \#4 for the $(5,1)$ TS and
\#2 and \#11 for the $(4,2)$ TS, are the unsatisfied check nodes. A
\textit{cycle} of length $c$ occurs when a path with $c$ edges
exists between a node and itself. There are no 4-cycles in this
code, but there are many 6-cycles within the subgraphs of Figure
\ref{fig: Error.TS_5_1}. The essential intuition pointing to these
as problematic for the decoder is that if $a$ bits have sufficient
noise to cause them to individually appear as soft 1's, while the
others in the subgraph are soft 0's, the check nodes with which the
$a$ variable nodes connect will be satisfied, tending to reinforce
the wrong state to the rest of the graph.  Only the unsatisfied
check(s) is a route through which messages come to reverse the
apparent (incorrect) state.

The number of edges within a TS, $|E_{TS}|$, is determined by the
number of variable nodes, $a$, and their degrees.

\begin{equation}
|E_{TS}| = \sum\limits_{i=1}^{a}d_{v_i}
\end{equation}

The number of check nodes participating in a dominant TS is
\textit{most often} given by

\begin{equation}
 |C_{TS}| = \frac{|E_{TS}| - b}{2} + b
\end{equation}

This equation assumes that all unsatisfied check (USC) nodes are
connected to one variable node in the TS and all satisfied checks
are connected to exactly two variable nodes from the TS. Subgraphs
with these properties are referred to as \textit{elementary trapping
sets} in the literature \cite{Milenkovic-05}. Since any odd number
of connections to TS bits will cause a check to be unsatisfied and
any even number will cause the check to be satisfied, it is not
guaranteed that all dominant TS are elementary. However, for
dominant TS, i.e. those with small $a$ and much smaller $b$, it is
evident that given $|E_{TS}|$ edges to spend in creating an $(a,b)$
TS, most edge permutations will produce $d_c = 2$ satisfied checks
and $d_c = 1$ USC's. Empirical evidence also supports this
observation as is seen in compiled tables of TS shown in Section
\ref{sec: decoder_search}.

\begin{figure}[h]
\begin{center}
\resizebox{3in}{3in}{\input{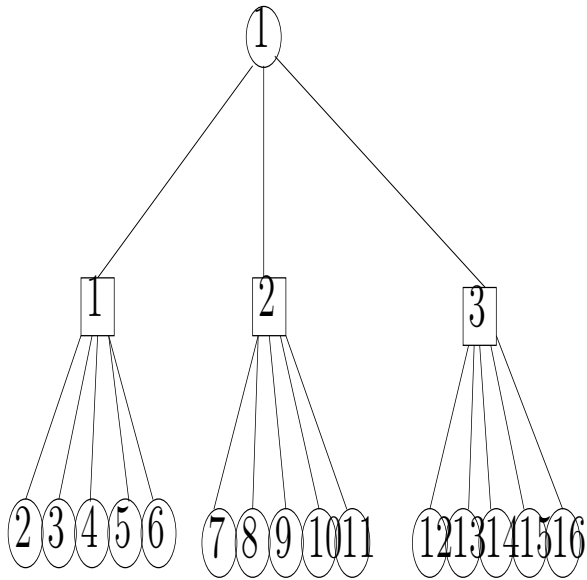}} \caption{Tree
showing first layer of check and variable nodes} \label{fig: tree}
\end{center}
\end{figure}

The \emph{girth}, $g$, of a graph is defined as the length of the
shortest cycle and we assume this value to be $g \geq 6$.  This
constraint is easily enforced when building low-density codes;
4-cycles are only present when two columns of ${\bf H}$ have 1's in
more than one common row. Now consider a tree obtained by traversing
the graph breadth-first from a given variable node. This tree has
alternating variable and check nodes in each tier of the tree.  For
this girth-constrained set of regular codes, a tree rooted at a
variable node will guarantee all $d_v$ nodes in the first tier of
variable nodes, in this case $d_v(d_c-1)=15$ nodes, will be distinct
as illustrated in Figure \ref{fig: tree}. If the root variable node
in the tree is set to a `1', then to satisfy all of the check nodes
in the first tier of check nodes, an odd number of variable nodes
under \textit{each} of the $d_c$ check nodes in the first tier of
variable nodes must be a `1.' Since with high probability the
dominant error events will correspond to elementary TS, we assume
that exactly one variable node associated with each check node in
the first tier is a `1.'  To enumerate all of these $1 + d_v =
4$-bit combinations we must consider all $(d_c-1)^{d_v}=125$
combinations in one tree, and then take all $n$ variable nodes as
the root of a tree, which entails $n(d_c-1)^{d_v}$ combinations for
a general regular graph. The search method of Section \ref{sec:
decoder_search} makes extensive use of these trees rooted at each
variable node.

\section{Trapping Set Search Method (Step 1)}  \label{sec: decoder_search}

The key to an efficient search for problematic trapping sets (and
low-weight codewords, which can be considered as an $(a,0)$ TS and
will thus further not be differentiated from other TS) lies in
significantly reducing the entire $n$-dimensional search space to
focus on the only regions which could contain the dominant error
events.  The low-density structure of Tanner graphs for LDPC codes
allows one to draw conclusions about the code's global behavior by
observing the local constraints within a few edges of each node.
This allows a search algorithm that searches the space local to each
variable node.

To see how the local graph structure can limit our search space for
dominant error events, consider two sets of length-$n$ bit vectors
with Hamming weight four. The first set, $S_1$, contains all such
vectors: $S_1 = \{\xv : w_H(\xv) = 4\}$. The second set, $S_2$,
consists of the constrained set of those four-bit combinations which
are contained within the union of a variable root node and one
variable node from \textit{each} of the three branches in the first
tier of variable nodes associated with that root node. When we
consider only satisfied check nodes connected to two active variable
nodes, the number of active check or variable nodes in tier $i$ is
$d_v(d_v-1)^{i-1}$. For example, the number of nodes in the first
tier of variable nodes $|V_1|$, and the first tier of check nodes
$|C_1|$, for $\{3,6\}$ codes, is $|V_1|=|C_1|=3$. Now assign the
variable nodes in the leftmost branch in the $i^{th}$ tier to set
$V_{i1}$ and label these from left to right for all variable node
sets in the $i^{th}$ tier. For example, in Figure \ref{fig: tree},
set $V_{11}$ contains nodes 2-6, set $V_{12}$ contains nodes 7-11,
and set $V_{13}$ contains nodes 12-16.  We do the same for check
node sets, and the first tier of check nodes would consist of only
the set $C_{11}$ containing checks 1-3.  Using this notation, set
$S_2$ can be defined as: $S_2 = \{\xv : (x_{j} =
1)\bigcap_{i=1}^{d_v}w_H(V_{1i})=1 \quad j = 1,\ldots,n \}$.

In a regular code there are $d_c-1$ elements satisfying
$w_H(V_{1i})=1$ for each $i = 1,\ldots,d_v$.  Since we choose these
elements from each of the $V_{1i}$ independently, there are
$(d_c-1)^{d_v}$ elements in $S_2$.  For a regular $\{3,6\}$ code,
$d_v + 1 = 4$, so the elements of $S_2$ are all 4-bit combinations,
and these combinations of variable nodes will ensure that at least
the three check nodes directly connected to the root variable node
are satisfied.  For example, if we choose the leftmost variable node
in each branch of Figure \ref{fig: tree}, a member of the set $S_2$
would be $\{v_{1}, v_{2}, v_{7}, v_{12}\}$.

Consider a (96,48) $\{3,6\}$ code where $|S_1| = \nchoosek{96}{4} =
3321960$ and $|S_2| = n(d_c-1)^{d_v} = 12000$.  This example
illustrates the large reduction in the number of vectors belonging
to $|S_2|$ as opposed to $|S_1|$, and thus results in a
correspondingly much smaller search space for dominant error events.
Notice the gap between the sizes of these two sets gets even larger
as block length increases.

The motivation for examining the smaller set of 4-bit combinations,
$|S_2|$, above was to limit the number of directions in
$n$-dimensional decoding space necessary to search for dominant
error events.  If a true ML decoder were available, a simple
technique can be utilized to find the minimum distance of a code
\cite{berrou-02}, which is similar to our problem of finding the
low-weight TS spectrum for a code. The idea is to introduce a very
unnatural noise, called an `error impulse,' in a single bit position
as input to the ML decoder. Unfortunately, the single-bit error
impulse method cannot be used with the MPA to find dominant error
events \cite{Hu-04} mainly because the MPA's objective is to perform
a \textit{bit} ML decision rule and not the \textit{vector} ML rule.
Typically when a large error impulse is input to one bit, the
decoder will correctly decode the error until the impulse reaches a
certain size where the channel input for that bit overrides the
$d_v$ check messages and flips that bit while leaving all $n-1$
other bits alone. Instead of the single-bit impulse, it makes more
sense to apply a multi-bit error with a smaller impulse magnitude in
each bit, as this would better simulate a typical high SNR noise
realization.

The choice of which bits to apply the impulse to is very important -
they should be a subset of the bits of a minimum distance TS. A good
candidate set of impulse bit locations is given by $S_2$. A
multi-bit error impulse should appear as a more `natural' noise to
the MPA and the 4-bit combinations of $S_2$ will be likely to get
error impulses into multiple bits of a dominant TS. For example,
suppose a minimum distance TS has a `1' in its first four bits. If
an error impulse $\epsilon_1$ were applied in all four of these
positions, leaving the other $n-4$ bits alone (i.e. the noise is
$-\epsilon_1[1,1,1,1,0,\ldots,0]$), would the message passing
algorithm decode to this minimum distance TS? It cannot be
guaranteed, but if the code block length is not too long, based on
extensive empirical evidence, the decoder will decode to this nearby
TS for sufficiently large $\epsilon_1$.  This MPA decoding behavior
leads to the following theorem.

\begin{theorem}
If $g \ge 6$, in a $\{3,6\}$-regular code, every $(a,b)$ TS with $a
> b$ must contain at least one 4-bit combination from $S_2$ among
the $a$ bits of the TS.
\end{theorem}

\begin{proof}
First notice that in a dominant TS, the TS variable nodes should not
be connected to more than one unsatisfied check (USC).  If a $d_v=3$
variable node were, then the 2 or 3 `good' messages coming to that
variable node should be enough to flip the bit, thus creating a
variable node that is connected to 0 or 1 USC's.  Thus, for a
dominant TS, we can assume that all variable nodes in the TS
connected to USC's are connected to only one USC.
  Now, take any variable node in the TS that is NOT connected to any USC's
(there will be $a-b$ of these) and use it as a root node to unroll
the graph to the first layer of variable nodes and notice that one
of these $(d_c-1)^{d_v}$ possible combinations of 4-bits (i.e. an
element of $S_2$) will all be within the $a$ TS bits.
\end{proof}

As block length $n$ increases, 4-bit impulses begin to behave like
the single-bit impulses described above, where there exists a
threshold $\epsilon_t$ such that for all impulse magnitudes below
this threshold the decoder corrects the message and if $\epsilon_1
> \epsilon_t$, then the decoder outputs a `1' in the four bits with the
impulse and sets the other $n-4$ bits to `0'.  For rate-1/2
$\{3,6\}$ codes this typically happens around $n=2000$.  One
modification to partially avoid this is to scale the other $n-4$
bits with another parameter, say $\gamma$.  In other words instead
of sending `1' in the $n-4$ noiseless bits, send $0<\gamma<1$.  This
allows the `bad' information from our four impulse bits to more
thoroughly propagate further out into the Tanner graph and
simultaneously lessens the magnitude of the `good' messages coming
in to correct the variable nodes where $\epsilon_1$ was applied.
This method also loses effectiveness as $n$ increases past 5000.

\begin{figure}[h]
\begin{center}

%\input{tree_impulse.pstex_t}
%\resizebox{10cm}{3cm} {\input{feedbackview.epic}}
%\resizebox{10cm}{7cm} {\input{tree_impulse.pstex_t}}

%\input{iterHdesign.pstex_t}
\resizebox{3.5in}{4in}{\input{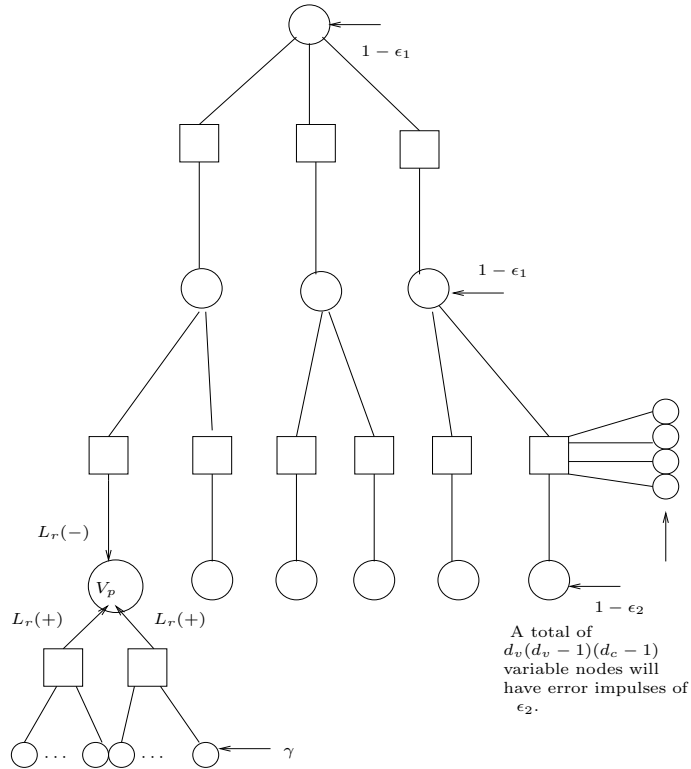}}

\caption{Tree for Deterministic Noise Input} \label{fig:err_impulse}
\end{center}
\end{figure}

For a still longer code, where say $g \ge 10$, the 4-bit impulse
method will generally fail unless modified again.  To see why,
consider the tree of Figure \ref{fig:err_impulse}, and we will use
the convention that the all-zeros message is sent and the LLR has
the probability of a bit being a zero in the numerator, thus a
`good' message which works to correct a variable node will have a
(+) sign and a `bad' message which works to reinforce the error
state will have a (-) sign.  For a $\{3,6\}$ code, the six variable
nodes at variable tier two of the tree ($V_p$ in the Figure) will
have four messages coming to them: the channel data $L_c (+)$, $L_r
(-)$ from checks connected to variable nodes which have an
$\epsilon_1$ error impulse input, and two $L_r (+)$ messages coming
from check nodes connected to noiseless variable nodes. Since the
minimum magnitude of the messages which are incoming to the three
check nodes neighboring one of the six $V_p$ variable nodes is equal
to $\gamma$, the three $L_r$ messages will have roughly the same
magnitude with belief propagation and exactly the same magnitude
with the min-sum algorithm. Thus, the two positive $L_r$ messages
overpower the single negative $L_r$ message, and the $L_c =
\gamma4E_s/N_o$ message provides even more positive weight to our
$LQ$ marginal probability calculation. To get around this problem
and force the decoder to return dominant error events, we find all
possible v-nodes that are at variable tier two and connected to the
$d_v(d_v-1)$ check nodes at check tier two of the tree. This will be
$d_v(d_v-1)(d_c-1) = 30$ variable nodes for a $\{3,6\}$ code. In
these positions input an error impulse with a smaller magnitude than
$\epsilon_1$, but larger than that of the outside parameter
$+\gamma$ values. Call this second error impulse value $\epsilon_2$.
This extra deterministic noise causes the negative $L_r$ messages
floating down to variable node tier three of the tree to have a much
larger magnitude than the positive $L_r$ messages coming up, and
this stronger `bad' information is more likely to cause the decoder
to fail on a dominant TS.

To recap, for a $\{3,6\}$ code, the deterministic input to the
decoder is now $[1-\epsilon_1, 1-\epsilon_1, 1-\epsilon_1,
1-\epsilon_1, 1-\epsilon_2, \ldots, 1-\epsilon_2, \gamma, \ldots,
\gamma]$.  It is important to have a definition of a TS which takes
into account the entire history of the decoding process and not just
the final state.  This new definition will eliminate ambiguities
that arise from the previous definition of a TS, which was based
solely on the combinatorial properties of a bit vector. Since the
entire decoding process involves many MPA iterations, a formal
definition is necessary to locate where in this dynamic process the
TS state was achieved. This definition will become important in
Section \ref{subsec:procedure_newerrors}.

\begin{definition} \label{Def:TS}
During the decoding process, a history of the hard decision, ${\bf
\hat{x}_l}$, of the message estimate must be saved at each iteration
$l$ and if the maximum number of iterations $I_{max}$ occurs and no
valid codeword has been found, the TS will be defined as the ${\bf
\hat{x}_l}$ which satisfies $\min\limits_l^{} w_H({\bf
\hat{x}_l}\Hm^T)$, where $l = 1,\ldots,I_{max}$.
\end{definition}

A practical example highlighting the power of this search method can
be seen with some long $\{3,6\}$ codes proposed by Takeshita
\cite{takeshita-05}.  In two rate-1/2 codes, with $n=8192$ and
$n=16384$, the search found many codewords at $d_H$ of 52 and 56
respectively.  This was a great improvement upon the results
returned from the Nearest-Nonzero Codeword Search of \cite{Hu-04}.
Table \ref{tab:good_err_impulse_example} gives the search parameters
required to find error events for some larger codes.  A total of 204
codewords with Hamming weight 24 were found in the Ramanujan
\cite{Rosenthal-00} code, 3775 codewords with Hamming weight 52 in
the Takeshita (8192,4096) code, and an estimated 4928 codewords with
Hamming weight 56 in the Takeshita (16384,8192) code. This last
estimate was determined by only searching the first 93 variable node
trees, which took 12 hours, and then multiplying the number of
codewords at Hamming weight 56 found up to that point (28 in this
case) by (16384/93). Thus the time of 2112 compute-hours, running on
an AMD Athlon 2.2 GHz 64-bit processor with 1 GByte RAM, is an
estimate\footnote{Throughout this paper, jobs requiring
compute-times larger than 8-10 hours have been executed on a Linux
cluster, where each node of the cluster has roughly the same
computing power as the aforementioned platform.  Thus, when large
compute-times are listed, this can be considered the number of
equivalent hours for a desktop computer.}. This reasoning would not
hold for most codes, but these algebraic constructions tend to have
a regularity about them from the perspective of each local variable
node tree. Although the large compute-times necessary for this
method may seem impractical, for codes of this length, a simple
Monte Carlo simulation would take much longer to find the error
floor and it would not collect the $d_{min}$ TS as this method does.

For the three example codes above, the listed Hamming weights are
believed to be their $d_{min}$.  There could be more codewords of
these Hamming weights, and it is possible that codewords of smaller
weight exist, but this is unlikely and the multiplicity of these
codewords found from our search is probably a tight lower bound on
the true multiplicity.  The argument here is the same used in
\cite{Hu-04}, only this method appears to be more efficient for most
types of low-density codes and has the advantage of also finding the
dominant TS which are usually the cause of an LDPC code error floor.

\begin{table}
  \centering
\begin{tabular}{|c||l|l|l|l|}
  \hline
%   &  &  &  &\\
     Code  & $\epsilon_1$ & $\epsilon_2$ & $\gamma$ & Time (Hrs)\\
  \hline
(4896,2448) (Ramanujan) & 5 & 2 & 0.4 & 24 \\
(8192,4096) (Takeshita) & 6.25 & 4 & 0.45 & 320 \\
(16384,8192)(Takeshita) & 6.25 & 5 & 0.4 & 2112* \\
  \hline
\end{tabular}
  \caption{Search Parameter Values, $E_b/N_o = 6$ dB, Max. 50 MPA Iterations}\label{tab:good_err_impulse_example}
\end{table}

Another way to deal with longer codes is to grow the tree one level
deeper, and apply the same $\epsilon_1$ to each of the variable
nodes at the root, variable tier one, and variable tier two in
Figure \ref{fig:err_impulse}.  For a $\{3,6\}$ code, this would give
sets of 10 bits in which to apply the error impulse.  The number of
these 10-bit combinations for each root node is
$(d_c-1)^{d_v}(d_c-1)^{d_v(d_v-1)} = 5^9 = 1953125$, which clearly
shows that this method is not nearly as efficient for long codes as
it is for short codes.  Still, the method will find an error floor,
or at least a lower bound on $P_f$, much faster than standard Monte
Carlo simulation.  It is possible to take variations of these sets
of bits; for example, if the code girth is at least 8, then all
6-bit combinations given by the root node, three variable nodes at
tier one and the first two variable nodes at tier two ($V_{21},
V_{22}$) are unique and make a good set of 6-bit error impulse
candidates. The number of bits needed to form an error impulse
capable of finding dominant TS is a function of $n,k$ and girth. For
larger $n$, more impulse bits are generally required.  This method
has been applied to many rate-1/2 codes with $n$ at least 10000, and
there was always a combination of parameters $\epsilon_1,
\epsilon_2$, and $\gamma$ which provided an enumeration of dominant
TS and codewords, leading to the calculation of the error floor in
much less time than what a standard Monte Carlo simulation would
require.

\subsection{Irregular Codes}  \label{subsec: decoder_search_irr}

Irregular codes can be constructed which require a lower $E_b/N_o$
to reach the `waterfall' threshold.  Unfortunately these codes often
suffer from higher error floors.  The new search technique can
efficiently determine what types of TS and codewords cause this bad
high-SNR performance.

The method of taking each of the $n$ variable nodes and growing a
tree from which we apply deterministic error impulses is the same.
The major observation is that nearly all dominant TS and codewords
in irregular codes contain most of their bits in the low-degree
variable nodes.  Most irregular code degree distributions contain
many $d_v = 2$ variable nodes, and these typically induce the low
error floor.  So, it makes sense to order the $n$ variable nodes
from smallest $d_v$ to largest and perform the search on the
smallest (i.e. $d_v = 2$) variable nodes first.  In fact, for all
irregular codes tested, the search for dominant TS can stop once
trees have been constructed for all of the variable nodes with the
smallest two $d_v$'s.  Note that the number of bits which receive an
error impulse is dependent on $d_v$ and will be $1+d_v$ if the tree
is only grown down to the first variable node tier.  The parameter
values for $\epsilon_1, \epsilon_2$, and $\gamma$ are also dependent
on the variable node degree of the root in a given tree. For
example, if the code has variable node degrees of [2 3 6 8], then
the associated $\gamma$'s might be [0.3 0.3 0.4 0.45], i.e. the
highest degree variable node of $d_v = 8$ would have an error
impulse in 9 bits, and thus it needs less help from the other $n-9$
bits to cause an error, so its $\gamma$ parameter can be set higher.

\subsection{High-Rate Codes}  \label{subsec: decoder_search_highrate}

LDPC codes of high rate contain check node degrees considerably
larger than their lower rate counterparts.  For example, in rate 0.8
regular $\{3,15\}$ codes, the 4-bit impulse method would require
$n(d_c-1)^{d_v} = n14^3 = 2744n$ decodings, much higher than the
$125n$ decodings for $\{3,6\}$ codes.  On a positive note, $n$ can
grow longer in these types of more densely-packed codes before the
search requires the help of the extra $\epsilon_2$ noise. The search
was applied to a group of codes proposed in \cite{Song-06} and
succeeded in locating dominant TS and codewords. One code had column
weights of 5 and 6 and row weights of 36. Instead of applying
$\epsilon_1$ to a variable node from each of the $V_{1i}$ sets in
variable tier one, which would require $d_v + 1$ impulse bits, we
instead pick a number less than $d_v$, call it $v_{num}$, in this
case 4, and choose all combinations of $v_{num}$ variable node sets
among the $V_{1i}$ sets.  Assuming the check node degrees are all
the same, the number of decodings $|D|$ required using this method
is given by \eqref{eq:decoder.v_num_sets}.

\begin{align} \label{eq:decoder.v_num_sets}
 |D| = \sum\limits_{i=1}^{|d_v|} |d_{v_{i}}|\nchoosek{d_{v_{i}}}{v_{num}}(d_c
 - 1)^{v_{num}}
\end{align}

\noindent where $|d_{v_{i}}|$ denotes the number of variable nodes
of degree $i$ and $|d_{v}|$ denotes the number of different variable
node degrees.

\subsection{Search Parameter Selection}\label{sec: experiment}

The choice of search parameters is very important in finding a
sufficient list of dominant error events in a reasonable amount of
compute time. The purpose of this section is to illustrate how the
search method depends on the magnitude of the error impulse,
$\epsilon_1$, for the simple 4-bit impulse.  Our example code, the
PEG $(1008, 504)$ $\{3,6\}$ code \cite{Mackay_codes} will require
$n(d_c-1)^{d_v} = (1008)5^3 = 126000$ total decodings in the search.
Each decoding will attempt to recover the all-zero's message from
the deterministic decoder input of $\gamma=0.6$ and $\epsilon_1$
varying over three values.  The SNR parameter will be $E_b/N_o = 6$
dB and a maximum of 50 BP iterations will be performed.  The error
impulse, $\epsilon_1$, will take on the values 3, 3.5, and 4.
Increasing $\epsilon_1$ increases the number of TS found while the
mean number of iterations required for each decoding is also
increased, which leads to the longer compute times needed to run the
search program for larger $\epsilon_1$. For example, if $\epsilon_1
= 3$, it might take 5 iterations on average to decode a message
block. If $\epsilon_1$ is increased to 4, it might take a mean of 10
iterations to decode, thus causing the program to take twice as
long, even though the total number of decodings, 126000, stays the
same.  The base case compute-time for this example code, with
$\epsilon_1 = 3$, is slightly under 40 minutes.

It appears that for most codes with $n < 2000$, there is an
$\epsilon_1$, call it  $\epsilon_1^*$, such that when  $\epsilon_1$
is increased above this level, few meaningful error events are
discovered beyond those which would be uncovered by using
$\epsilon_1^*$. So, by determining the probability of frame error
contributed by those events found by running the search program with
$\epsilon_1^*$, we should have a reasonably tight lower bound on
$P_f$ for the code.  How do we best find this $\epsilon_1^*$? There
is probably no practical analytical solution to this question, but
Tables \ref{tab:H1008PEG_TS_1} \ref{tab:H1008PEG_TS_2} and
\ref{tab:H1008PEG_TS_3} list the error events returned from the
search using three different values of $\epsilon_1$ and will help
illustrate the issue. The columns in the tables, from left to right,
represent the TS class, multiplicity of that class,
squared-Euclidean distance to the error threshold found by a
deterministic noise directed towards the TS (averaged over each
member of a specific TS class, this error threshold will be
explained in Section \ref{sec: error_boundary}), and the number of
TS from this class that are \textit{elementary}
\cite{Milenkovic-05}, meaning all unsatisfied checks have one edge
connected to the TS bits.

\begin{table}
  \centering
\begin{tabular}{|c||r|r@{.}l|r|}
  \hline
%   &  &  & \\
%   \multicolumn{5}{|c|}{Error Class}\\
   Error Class  &  Multiplicity &  \multicolumn{2}{|c|}{$d_{\E}^2$} &
$|TS|_{Elem}$ \\
%   $|TS|_{Elem}$}\\
%     Error Class  &   Multiplicity &  $D_{\E}$ & & $|TS|_{Elem}$ \\
  \hline
(6,2) & 5 & 12&47 & 5\\
(4,2) & 6 & 12&43 & 6\\
(8,2) & 3 & 12&79 & 3\\
(10,2) & 3 & 14&29 & 3\\
(9,3) & 27 & 19&93 & 27\\
(7,3) & 57 & 22&27 & 57\\
(12,2) & 1 & 19&61 & 1\\
(5,3) & 21 & 29&32 & 21\\
(11,3) & 2 & 79&34 & 2\\
(10,4) & 5 & 70&70 & 5\\
(8,4) & 8 & 77&49 & 7\\
(6,4) & 21 & 55&89 & 21\\
  \hline
\end{tabular}
  \caption{Dominant Error Event Table for (1008,504) PEG - $E_b/N_o=6$ dB, $\epsilon_1 = 3.0$, $\gamma = 0.6$, 50 iterations}\label{tab:H1008PEG_TS_1}
\end{table}

\begin{table}
  \centering
\begin{tabular}{|c||r|r@{.}l|r|}
  \hline
%   &  &  & \\
   Error Class  &  Multiplicity &  \multicolumn{2}{|@{}c@{}|}{$d_{\E}^2$} &
$|TS|_{Elem}$ \\
  \hline
(6,2) & 5 & 12&47 & 5\\
(4,2) & 6 & 12&43 & 6\\
(8,2) & 3 & 12&79 & 3\\
(10,2) & 4 & 14&26 & 4\\
(12,2) & 4 & 16&08 & 4\\
(9,3) & 89 & 22&20 & 88\\
(7,3) & 97 & 23&58 & 97\\
(14,2) & 2 & 17&74 & 2\\
(11,3) & 8 & 39&39 & 8\\
(5,3) & 90 & 33&90 & 90\\
(10,4) & 30 & 64&40 & 29\\
(8,4) & 82 & 56&71 & 82\\
(6,4) & 127 & 55&57 & 127\\
(12,4) & 2 & 45&85 & 0\\
(7,5) & 15 & 82&75 & 15\\
(9,5) & 4 & 98&12 & 4\\
(11,5) & 5 & 120&52 & 4\\
(13,3) & 1 & 159&03 & 0\\
(15,3) & 1 & 183&49 & 1\\
  \hline
\end{tabular}
  \caption{Dominant Error Event Table for (1008,504) PEG - $E_b/N_o=6$ dB, $\epsilon_1 = 3.5$, $\gamma = 0.6$, 50 iterations}\label{tab:H1008PEG_TS_2}
\end{table}

\begin{table}
  \centering
\begin{tabular}{|c||r|r@{.}l|r|}
  \hline
 %  &  &  & \\
   Error Class  &   Multiplicity &  \multicolumn{2}{|@{}c@{}|}{$d_{\E}^2$} &
 $|TS|_{Elem}$ \\
  \hline
(6,2) & 5 & 12&47 & 5\\
(4,2) & 6 & 12&43 & 6\\
(8,2) & 3 & 12&79 & 3\\
(10,2) & 4 & 14&26 & 4\\
(12,2) & 6 & 16&64 & 6\\
(9,3) & 113 & 21&29 & 111\\
(7,3) & 110 & 26&59 & 109\\
(14,2) & 3 & 17&73 & 3\\
(11,3) & 49 & 23&38 & 48\\
(16,2) & 1 & 19&36 & 1\\
(13,3) & 9 & 52&08 & 9\\
(5,3) & 104 & 35&88 & 104\\
(10,4) & 124 & 48&36 & 116\\
(8,4) & 469 & 45&96 & 466\\
(12,4) & 12 & 94&28 & 5\\
(6,4) & 384 & 54&95 & 383\\
(9,5) & 104 & 90&18 & 100\\
(11,5) & 10 & 100&19 & 7\\
(14,4) & 3 & 97&89 & 1\\
(7,5) & 176 & 80&71 & 172\\
(15,3) & 3 & 124&18 & 0\\
(13,5) & 1 & 159&03 & 1\\
(15,5) & 1 & 183&49 & 0\\
  \hline
\end{tabular}
  \caption{Dominant Error Event Table for (1008,504) PEG - $E_b/N_o=6$ dB, $\epsilon_1 = 4.0$, $\gamma = 0.6$, 50 iterations}\label{tab:H1008PEG_TS_3}
\end{table}

The major point to observe from the tables is that the first three
rows, representing the most dominant error events, for this code TS
of classes $(6,2)$, $(4,2)$, and $(8,2)$, are unchanged for each of
the four search executions.  This robustness to an uncertain
$\epsilon_1^*$ is important for this method to be a viable solution,
since $\epsilon_1^*$ will have to be iteratively estimated for a
given code.  Extensive Monte Carlo simulations with the nominal
noise density in the higher SNR region verify that indeed the first
three rows include the error events most likely to occur.

\section{Locating TS Error Boundary (Step 2)}  \label{sec: error_boundary}

Once a list of potential dominant error events has been compiled, it
is not a simple task to determine which of these bit vectors will
cause the decoder the most trouble.  For an ML decoder this is not
an issue because the Hamming weight, $w_H$, is enough information to
determine the two-codeword error probability: $P(\xv_1 \rightarrow
\xv_2) = Q(\sqrt{2w_H E_s/N_o})$.  For a TS, it is possible to
determine the error contribution of a certain bit vector by either
conditioning the probability of error on the magnitude of the noise
in the direction of the TS bits \cite{richardson} or by using a
mean-shifting IS procedure \cite{cavus-05}.  Both of these methods
require a simulation with at least a few thousand noisy messages per
SNR to get an accurate measurement of the $P_f$ contributed by the
TS in question.  The idea proposed here is to send a deterministic
noise in the direction of the TS bits and let the decoder tell us
the magnitude of noise necessary to cross from the correct decoding
region into the error region and use this information to quickly
determine \textit{relative} error performance between different TS,
not necessarily of different $(a,b)$ type.  This method only maps
out the point of the error boundary which is along the line
connecting the all-ones point in $n$-dimensional signal space with
the point on the $n$-dimensional hypercube associated with the given
TS.  It requires only $p$ decodings to find this point of the error
boundary with accuracy $(l_{max} - l_{min})/2^{p}$ since a binary
search, as described below, has complexity $O(\log L)$, where
$L=2^{p}$ is the number of quantization bins in between $l_{max}$,
the largest magnitude in a dimension where a TS bit resides, and
$l_{min} = 1$, which would be on the error boundary if the TS were
an actual codeword. $l_{max}=3.5$ is the value used in this
research.  The procedure to locate the error boundary is to first
input the vector $\yv = [1-I(1)\epsilon, 1-I(2)\epsilon, \ldots,
1-I(n)\epsilon]$ to the decoder, where $I(i)$ is the indicator
function for whether the $i^{th}$ bit belongs to the TS and the
magnitude of $\epsilon$, call it $\epsilon_1$, is $(l_{min} +
l_{max})/2$. If the decoder \textit{corrects} this deterministic
error input, then for the second iteration, and in general for the
$i^{th}$ iteration, update $\epsilon_i = \epsilon_{i-1} + (l_{max} -
l_{min})/2^{i}$ and apply this to the decoder input. If the first
input vector resulted in an \textit{error}, then set $\epsilon_i =
\epsilon_{i-1} - (l_{max} - l_{min})/2^{i}$. This process will be
repeated $p$ times, which tells us to within $(l_{max} -
l_{min})/2^{p}$ how close the error boundary is, requiring only $p$
decodings.  $p = 10$ is used in this research and should be more
than adequate for most purposes.

\begin{figure}[htbp]
\begin{center}
\resizebox{3.0in}{3in}{\input{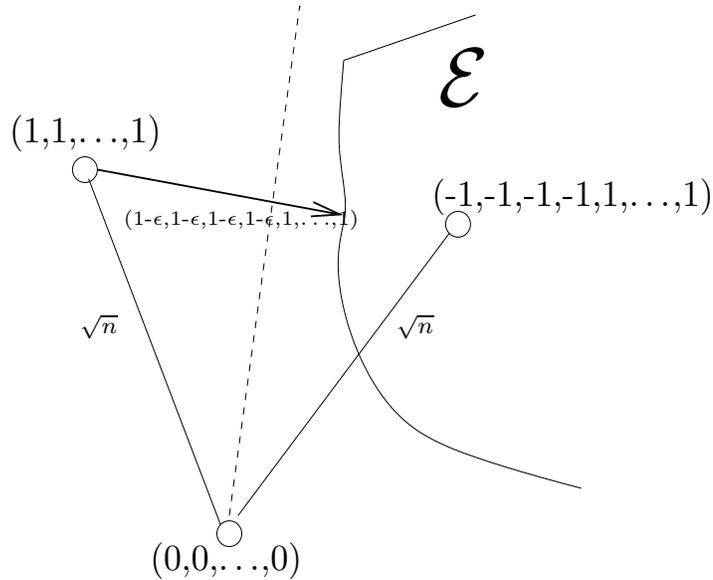}}
\caption{Deterministic Error in TS Direction}
\label{fig:errorboundary}
\end{center}
\end{figure}

All non-codeword TS, since they have at least one unsatisfied check
(USC), should have a distance to the error boundary that is at least
as large as the distance to a $w_H = a$ codeword error boundary.
Finding the error contribution of a specific TS is analogous to the
two-codeword problem, except the error region is much more
complicated than the half-space decision region resulting from the
two-codeword problem. The situation is depicted in Figure
\ref{fig:errorboundary} where we assume a $(4,2)$ TS exists among
the first 4 bits of the $n$-length bit vector. Consider the $n-1$
dimensional plane bisecting the line joining the {\bf 1} vector
(all-zeros codeword) and the (-1,-1,-1,-1,1,...,1) TS; this plane
would represent a half-space boundary if the TS were actually a
codeword and the decoder were ML. The arrow starting at the point
{\bf 1} (the signal space coordinates of codeword {\bf 0}) and
directed towards the TS shows where the error region begins. The
shape of the error region is very complicated and this
two-dimensional figure does not accurately represent its true shape,
but it makes intuitive sense that the nearest point in the error
region should be in the direction of the bits involved in the TS.

%\begin{figure} [h]
%\begin{center}
%\epsfig{file=plots/TSerror, height=4.0in, width=3.2in}
%\caption{Codeword/TS Error Decision Region for (4,2)
%TS}\label{fig:TSerror}
%\end{center}
%\end{figure}

Table \ref{tab:H504_TS} lists the dominant TS found with the search
for a (504,252) regular $\{3,6\}$ code with girth eight
\cite{cole-06}.  The parameters of the search were set as follows:
$\epsilon_1 = 3.6, \gamma = 0.8$, 50 iterations, $E_b/N_o = 5$ dB.
The column labeled $d_{\E}^2$ denotes the \textit{average}
Euclidean-squared distance to the error boundary for the TS class in
that row.  For example, if a deterministic noise impulse with
magnitude $\epsilon$ were applied in each of the $a$ bits of an
$(a,b)=(10,2)$ TS and the decoder switched from correctly decoding
to an error at $\epsilon = 1.5$, then $d_{\E}^2 = a\epsilon^2 =
10(1.5)^2 = 22.5$.  The rows are ordered by dominance, where the
\textit{minimum} $d_{\E}^2$ among all TS of a class determines
dominance. For example, the (11,3) TS on average have a larger
$d_{\E}^2$ (23.3 in this case) than the (9,3) TS, which has an
average of 20.7, but the minimum $d_{\E}^2$ among the 20 (11,3) TS
is less than the minimum among the 186 (9,3) TS.  This behavior is
in contrast to valid codewords, where all codewords with a given
$w_H$ will have the same error contribution in the two-codeword
problem.  The $(10,2)$ TS class had a member with the smallest
$d_{\E}^2$ among all TS found for this code.  Knowing $d_{\E}^2$
does not tell us exactly what contribution a TS gives to $P_f$, but
$Q(\sqrt{2d_{\E}^2E_s/N_o})$ does approximate this contribution to
within a couple orders of magnitude.

\begin{table}
  \centering
\begin{tabular}{|c||r|r@{.}l|r|}
  \hline
%   &  &  & \\
   Error Class  &   Multiplicity &  \multicolumn{2}{|@{}c@{}|}{$d_{\E}^2$} &
 $|TS|_{Elem}$ \\
  \hline
(10,2) & 22 & 15&01 & 22\\
(12,2) & 5 & 15&87 & 5\\
(11,3) & 20 & 23&29 & 17\\
(9,3) & 186 & 20&68 & 186\\
(10,4) & 46 & 28&64 & 40\\
(13,3) & 4 & 19&76 & 3\\
(8,4) & 303 & 27&78 & 300\\
(7,3) & 106 & 26&09 & 106\\
(12,4) & 3 & 41&87 & 2\\
(6,4) & 1178 & 41&14 & 1178\\
(9,5) & 15 & 27&99 & 13\\
(7,5) & 41 & 32&92 & 41\\
(11,5) & 1 & 26&82 & 0\\
(5,3) & 24 & 41&30 & 24\\
  \hline
\end{tabular}
  \caption{Dominant Error Event Table for (504,252) $\{3,6\}$ Code - $E_b/N_o=6$ dB, $\epsilon_1 = 3.5$, $\gamma = 0.6$, 50 iterations}\label{tab:H504_TS}
\end{table}

One way to get more confidence in the validity of using $d_{\E}^2$
as a criteria to establish which TS are most dominant is to simulate
a large number of noisy message frames using the nominal Gaussian
noise density at a higher SNR and tabulating which TS classes these
errors fall into.  This is done for the PEG (1008,504) code at
$E_b/N_o = 4.0$ dB and the results are in direct correlation with
what is expected from the $d_{\E}^2$ values computed with our
deterministic noise impulse (See Table \ref{tab:H1008PEG_TS_3} for
dominant TS of this code). This code has five (6,2) TS which
dominate and two of them have a $d_{\E}^2$ smaller than the rest (at
4.0 dB), and these specific TS are indeed much more likely to occur
with the nominal noise density. Two (8,2) TS are also very dominant.
Table \ref{tab:H1008PEG_TS} shows all of the TS which had more than
one error with the nominal noise density.  $6(10)^8$ trials at
$E_b/N_o = 4.0$ dB were performed and a total of 85 errors were
recorded. The value in column two of Table \ref{tab:H1008PEG_TS}
denotes the number of times the five most dominant error events
occur. All but five of these 85 total errors were from TS that were
represented in the list in Table \ref{tab:H1008PEG_TS_3} obtained
with the new search method.

\begin{table}
  \centering
\begin{tabular}{|c||l|l|}
  \hline
 %  &  &  \\
     TS  & $\# Errors$ &  $d_{\E}^2$ at 4.0 dB\\
  \hline
$(6,2)_1$ & 28 & 14.31 \\
$(6,2)_2$ & 20 & 14.36 \\
$(8,2)_1$ & 11 & 14.35 \\
$(8,2)_2$ & 8 & 14.46 \\
$(10,2)_1$ & 3 & 15.54 \\
%  \hline
%MS & $(12,2)_1$ &  & \\
%MS & $(12,2)_2$ &  & \\
  \hline
\end{tabular}
  \caption{Monte Carlo Verification of $d_{\E}^2$}\label{tab:H1008PEG_TS}
\end{table}

%(6,2)_1 =  71   287   431   517   854   938
%(6,2)_2 =  99   330   419   595   697   885
%(8,2)_1 =  19   130   193   208   290   535   744   771
%(8,2)_2 =  62   200   221   322   436   511   905   972
%(10,2)_1 =   77   236   240   275   289   333   469   965   980   986

\section{Importance Sampling (IS) (Step 3)}  \label{sec: IS}

In IS, we statistically bias the received realizations in a manner
that produces more errors \cite{Srinivasan-02, Smith-97, wu-95}.
Instead of incrementing by one for each error event ($I_{e}$), as
for a traditional Monte Carlo (MC) simulation, a `weight' is
accumulated for each error to restore an unbiased estimate of
$P_{f}$. This strategy, if done correctly, will lead to a greatly
reduced variance of the estimate compared to standard MC.

$f^{*}(\yv)$ denotes the (IS) biasing density and it is incorporated
into the MC estimate as follows:

\begin{align}
P_{f}       &  \triangleq E[I_{e}(\yv)] \notag\\
            &  = \int_{\Re^n} I_{e}(\yv)f(\yv)d\yv  \notag\\
            &  = \int_{\Re^n} I_{e}(\yv)\frac{f(\yv)}{f^{*}(\yv)}f^{*}(\yv)d\yv \notag \\
            &  = E_{*}[I_{e}(\yv)w(\yv)] \notag
\end{align}

This gives an alternate sampling estimator

\begin{align} \label{eq: IS.IS intro}
     \hat{P}_{f_{IS}}  &  = \frac{1}{L}\sum\limits_{l=1}^{L}I_{e}(\yv_{l})w(\yv_{l})
\end{align}

$L$ realizations $\yv_{l}$ are generated according to $f^{*}(\yv)$,
the biased density.  If $\yv_{l}$ lands in the error region then the
weight function, $w(\yv_{l}) = \frac{f(\yv_{l})}{f^{*}(\yv_{l})}$,
is accumulated to find the estimate of $P_{f}$.  MC can be seen as a
special case of this more general procedure, with $f^{*}(\yv) =
f(\yv)$.  $\hat{P}_{f_{IS}}$ is unbiased and has a variance given by

\begin{align} \label{eq: IS.IS var}
 Var[\hat{P}_{f_{IS}}]  &  = E_{*}[\frac{1}{L^{2}}(\sum\limits_{l=1}^{L}
I_{e}(\yv_{l})w(\yv_{l}))^{2}] - P_{f}^{2} \notag\\
            &  = \frac{1}{L^{2}}(LE_{*}[I_{e}^{2}(\yv)w^{2}(\yv)] + L(L-1)P_{f}^{2})  - P_{f}^{2} \notag\\
            &  = \frac{E_{*}[I_{e}(\yv)w^{2}(\yv)] - P_{f}^{2}}{L}\notag \\
            &  = \frac{E[I_{e}(\yv)w(\yv)] - P_{f}^{2}}{L} \notag\\
            &  = \frac{\int\limits_{\mathcal{E}}^{}w(\yv)f(\yv)d\yv - P_{f}^{2}}{L}
\end{align}

Notice that because $f^{*}(\yv)$ is in the denominator of $w(\yv)$,
it must be non-zero over the error region $\mathcal{E}$, else
$Var[\hat{P}_{f_{IS}}]$ is unbounded. The key quantity here is the
first term on the RHS of the last line in (\ref{eq: IS.IS var}). For
the IS method to offer a smaller variance than MC, this quantity
must be less than $P_{f}$ which appears as the first term on the RHS
of $Var[\hat{P}_{f_{MC}}]$. We will denote this term as $V$ and
estimate it on-line using MC, where the samples are taken from
$f^{*}(\yv)$.

\begin{equation} \label{eq: IS.IS var estimate}
\begin{tabular} {ll}
 $ \hat{V} $ &  $= \frac{1}{L} \sum\limits_{l = 1}^{L}w(\yv_l)^{2} $ \\
\end{tabular}
\end{equation}

Both $\hat{V}$ and $\hat{P}_{f_{IS}}$ rely on the same samples,
$\yv_l$, and this circular dependence means that if we are
underestimating $P_{f}$, then we are likely underestimating $V$.
$\hat{V}$ can give us some confidence in $\hat{P}_{f_{IS}}$, but the
simulator must make sure that $\hat{P}_{f_{IS}}$ passes several
consistency checks first.  It will most often be the case that
$\hat{P}_{f}$ is being underestimated. One check is employing the
sphere packing bound \cite{shannon-59}, which can be used as a very
loose lower bound that no code could possibly exceed.  One benefit
of tracking $\hat{V}$ is that if it indicates a poor estimate, then
$\hat{P}_{f}$ is definitely not accurate, but the converse is not
true.

Using an $f^{*}$ with the same properties as the nominal $f$, except
for a shifted mean to center the new density at the error boundary,
has been shown to provide the smallest $Var[\hat{P}_{f_{IS}}]$ in
the two-codeword problem \cite{wu-95, Srinivasan-02} and large
deviations theory \cite{bucklewbook-04} also suggests that
mean-shifting to the nearest error regions in $n$-dimensional signal
space should provide the optimal $f^{*}$. The $f^{*}$ proposed for
determining error performance of LDPC codes is based on a weighted
sum of mean-shifted $f^{*}_i$ densities, where there are $M$ nearby
error events used to form $f^{*}$.

\begin{align} \label{eq: f_star}
    f^*(\yv) & = \frac{1}{M}\sum_{i=1}^M
    \frac{1}{(2\pi\sigma^2)^{n/2}}\exp-\frac{|\yv - {\bf
    \mu_i}|^2}{2\sigma^2}
\end{align}

The ${\bf \mu_i}$ are $n$-bit vectors with zeros in all places
except for ones in the $a$ bits of an $(a,b)$ dominant TS or the
$w_H$ bits of a low-weight codeword.  The choice of using a
magnitude of `1' in the $a$ TS bit positions is probably not the
most efficient. When shifting towards valid codewords, `1' is the
optimal value for this mean-shifted $f^*$ \cite{wu-95}.  The
distance to the error boundary for a TS, as found in step two of our
procedure, always has a magnitude $\geq1$ in the $a$ bit positions
of the TS. So, if mean-shifting to the \textit{boundary} of the
error region is the most efficient IS procedure, then this shift
value should be used instead of `1'.  Since the two-codeword error
regions of TS are not in the shape of a half-space, this argument is
not quite correct, so although a more efficient simulation can be
performed by increasing the shift value to have a magnitude larger
than one, care must be used to not \textit{over} bias the shift
point, which will return a $\hat{P}_{f_{IS}}$ which is too small
\cite{Smith-97}. This weighted-sum IS density should catch many of
the `inbred' TS which were not explicitly caught in the initial TS
search phase, but which share many bits with the TS vectors that
were found and thus are `close by' in $n$-dimensional decoding
space.

The weighted-sum $f^{*}$ IS density should work best at moderate SNR
where many error events contribute to the error floor. At the
highest SNR's of interest, large deviations theory suggests that
only the nearest error events in $n$-dimensional decoding space
contribute to $P_f$ \cite{Sadowsky-90}.  If there are a small number
of these nearest events, then it is appropriate to break
$\mathcal{E}$ up into $\alpha$ regions $\mathcal{E}_i, i =
1,\ldots,\alpha$ corresponding to each of the minimum distance error
events. $\alpha$ is the total number of these minimum distance
events. An example of this technique used on a $\{4,8\}$ code will
be given in Section \ref{sec: results}.

\section{Error Floor Estimation Procedure}  \label{sec: procedure}

Our procedure consists of three steps which all make use of the
decoding algorithm to map out the $n$-dimensional region of
$\mathcal{E}$ and estimate $P_f$.  Since this region is very
dependent on the particular decoding algorithm and its actual
implementation details, e.g. fixed-point or full double-precision
values for messages, it is imperative to make use of the specific
decoder to determine $\mathcal{E}$.

The first step uses the search method proposed in \cite{cole-06} and
expanded upon in Section \ref{sec: decoder_search} to obtain a list
of dominate error events. This list is dependent on the decoding
algorithm since certain TS might be more problematic for say the
full belief propagation implementation as opposed to the min-sum
approximation.  The size of the list is also a function of the
parameters $\epsilon_1, \epsilon_2, \gamma$, and $E_b/N_o$.  It is
important to choose these parameters to find all of the dominant
error events, while still keeping the average number of iterations
per decoding as small as possible to avoid wasting computing time.
We will see in some examples that as long as all of the minimum
distance error events are included in this list, it is possible to
miss some of the moderately dominant error events and still get an
accurate estimate of $P_f$.  This is an important benefit over the
method which breaks up the error region into separate pieces for
each possible bit vector as proposed in \cite{richardson, cavus-05},
where the estimate of $P_f$ will almost assuredly be below the true
value, as it is nearly impossible to guarantee all important error
events have been accounted for, especially for longer and higher
rate codes.

\begin{figure}[h]
\begin{center}

\resizebox{3in}{3in}{\input{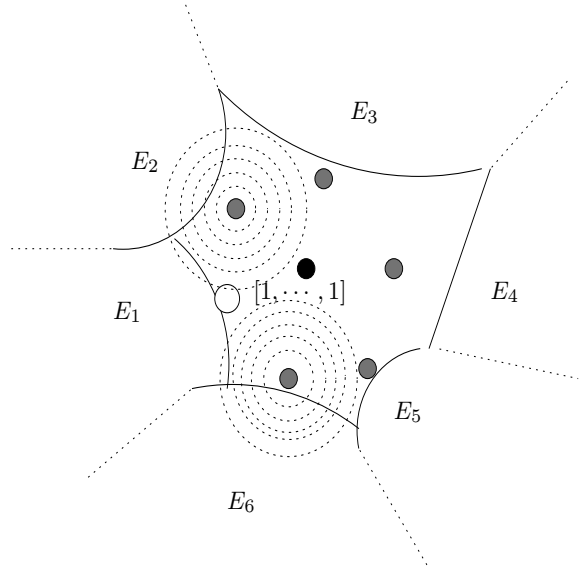}}

\caption{IS mean-shifting to cover error region}
\label{fig:IS_error_region}
\end{center}
\end{figure}

To expand on this, consider a simple toy example where there are six
dominant error regions and our initial list contains all but one of
these. Figure \ref{fig:IS_error_region} shows the error region
surrounding the all-zeros codeword.  The small, grey filled circles
between the all-ones point in $n$-dimensional space and the nearest
error regions are the mean-shift points in $f^*$. The single
white-filled circle in front of $\E_1$ represents the dominant error
event which was not accounted for in the initial list obtained with
the search from step one of the three-step procedure. Notice that
the two mean-shift points towards the regions $\E_2$ and $\E_6$ are
`close enough' in $n$-dimensional space to have a high probability
of landing some $f^*_2$ or $f^*_6$ noise realizations in $\E_1$.
This will allow the $\E_1$ contribution to be included in the total
$P_f$ estimate. If, on the other hand, there are $\E_i$ such that no
mean-shift points are near enough to these regions to have
significant probability of producing noise realizations in them,
then $\hat{P}_{f_{IS}}$ will with high probability (essentially the
same probability of not getting a hit in $\E_i$) underestimate $P_f$
by the amount of the error probability that lies in $\E_i$. So the
paradox is that even though $\hat{P}_{f_{IS}}$ is unbiased, it can
with high probability underestimate the true $P_f$ when using this
particular $f^*$.

To form an $f^*$ which adequately covers the error region without
needlessly including too many shift points which are unlikely to
offer any error region hits and only serve to complicate $f^*$, we
make use of the $d_{\E}^2$ values returned from stage two.  The
third column of Table \ref{tab:H504_TS} lists the $d_{\E}^2$ values
for a (504,252) $\{3,6\}$ girth eight LDPC code \cite{cole-06}. From
this column, it can be seen that the (10,2) and (12,2) TS are the
most dominant error events.  The (6,4) TS have a very large
multiplicity of at least 1178, but with an average $d_{\E}^2$ of
41.14, they don't contribute much to $P_f$ at higher SNR. Still,
there are some (6,4) TS with much smaller $d_{\E}^2$ than the
average over this class, so some of these should be included in
$f^*$. Thus, a good strategy would be to order the entire list
provided by the search and pick the $M$ TS with the smallest
$d_{\E}^2$ to include in $f^*$ for the third step of the procedure.
$M$ will be based on the parameters $n$ and $k$ of the code as well
as the size of the initial TS list. A larger $M$ will provide a more
accurate estimate of $P_f$, but will require more total noisy
message decodings if we use a fixed number of decodings $P$ for each
mean-shifted $f^*_i$.

The final step of our procedure takes the $M$ error events with the
smallest $d_{\E}^2$ and forms a weighted sum of mean-shift $f^*_i$
densities for $f^*$ \eqref{eq: f_star}. Equation \eqref{eq:
modulo_f_star} below confirms that deterministically generating $P$
realizations for each of our $M$ $f^*_i$ densities will form a
valid, unbiased, $\hat{P}_{f_{IS}}$.

\begin{align} \label{eq: modulo_f_star}
 \hat{P}_{f_{IS}}  &  = \frac{1}{M}\sum\limits_{m=1}^{M}
\frac{1}{P}\sum\limits_{l=1}^{P}
\frac{I_{e}(\yv)f(\yv)}{\frac{1}{M}\sum\limits_{p=1}^{M}f^*_{p}(\yv)}
\end{align}
The expected value is
\begin{align*}
 E[\hat{P}_{f_{IS}}]  &  = \frac{1}{M}\sum\limits_{m=1}^{M}
E[\frac{I_{e}(\yv)f(\yv)}{\frac{1}{M}\sum\limits_{p=1}^{M}f^*_{p}(\yv)}]
 = \frac{1}{M}\sum\limits_{m=1}^{M}
\int\limits_{\mathcal{E}}^{}\frac{f(\yv)f^*_m(\yv)}{\frac{1}{M}\sum\limits_{p=1}^{M}f^*_{p}(\yv)}d\yv\\
 &  = \frac{1}{M} \int\limits_{\mathcal{E}}^{}\frac{f(\yv)\sum\limits_{m=1}^{M}f^*_m(\yv)}{\frac{1}{M}\sum\limits_{p=1}^{M}f^*_{p}(\yv)}d\yv
   = \int\limits_{\mathcal{E}}^{}f(\yv)d\yv
   = P_f
\end{align*}

One implementation issue which must be addressed concerns the
numerical accuracy of the weight function calculation.  Any computer
will evaluate $e^x = 0$ when $x < -N$ for some positive $N$.  When
the block length is large or the SNR is high, all of the terms in
$f^*$ could equate to zero, giving a weight of $\frac{x}{0} =
\infty$.  To avoid this, a very small constant term can be added in
both the $f(\yv)$ and $f^*(\yv)$ distributions which will, with high
probability, ensure that $|x| < N$ while not affecting the value of
the weight function.  The constant term $\psi$ is added in the
second line of the following equation:

\begin{align} \label{eq: scale_factor}
w(\yv)  &  =
\frac{(2\pi\sigma^2)^{n/2}}{(2\pi\sigma^2)^{n/2}}\frac{\exp-\frac{(\yv
- {\bf 1})^2}{2\sigma^2}} {\frac{1}{M} \sum\limits_{m=1}^{M}
\exp-\frac{(\yv - {\bf 1} + {\bf \mu_m})^2}{2\sigma^2}} =
\frac{\exp(-\frac{(\yv - {\bf 1})^2}{2\sigma^2}+\psi)} {\frac{1}{M}
\sum\limits_{m=1}^{M} \exp(-\frac{(\yv - {\bf 1} +
{\bf \mu_m})^2}{2\sigma^2}+\psi)}\notag\\
 & = \frac{\exp\psi\exp-\frac{(\yv -
{\bf 1})^2}{2\sigma^2}} {\exp\psi\frac{1}{M}\sum\limits_{m=1}^{M}
\exp-\frac{(\yv - {\bf 1} + {\bf \mu_m})^2}{2\sigma^2}}  =
\frac{\exp-\frac{(\yv - {\bf 1})^2}{2\sigma^2}}{\frac{1}{M}
\sum\limits_{m=1}^{M} \exp-\frac{(\yv - {\bf 1} + {\bf
\mu_m})^2}{2\sigma^2}}
\end{align}

If $\psi$ is chosen to be $\frac{n}{2}$, then the argument in the
exponent of the term in the denominator corresponding to the $f^*_i$
centered about the $i^{th}$ shift point will be $ -(\frac{\sum_{i =
1}^{n} (n_i)^2}{2\sigma^2} - \frac{n}{2})$.  Since the $z =
\frac{\sum_{i = 1}^{n} (n_i)^2}{2\sigma^2}$ term is a scaled
$\chi^2$ random variable with $E[z] = n\frac{\sigma^2}{2\sigma^2}
=\frac{n}{2}$, all $M$ terms in the denominator of the weight
function will be zero only if all of the $\chi^2$ random variables
fall more than $N$ from their mean, which is highly unlikely,
especially for large $n$ and high SNR.

\subsection{New Error Events}  \label{subsec:procedure_newerrors}

One way to gain insight into how well the initial list contains
important error events is to keep track of all errors which occur
during the IS simulation, and call these `hits.' When a hit occurs,
we see if this TS is the same as the bit string towards which we
biased for that noise realization. If it is, this will be called an
`intended hit.' As SNR increases, the number of intended hits should
approach the number of hits, because the noise `clouds' are more
concentrated and less likely to stray from the error region near the
shift point. If the decoder does not converge to the all-zeros
codeword after the maximum number of iterations and a new TS, as
defined by Definition \ref{Def:TS}, occurs during the decoding
process which is not among the initial $M$ shift points, then add
this new error event to a cumulative list. Continue making this list
of new error events and their frequency of occurrence. The list of
new error events will gauge how thorough the initial list of TS
covers the error region. Because noise realizations have occurred in
the new error regions associated with the new error events, these
regions are effectively considered in $\hat{P}_{f_{IS}}$, even
though they weren't included in the $M$ shift-points ahead of time.
The new TS should contain many bits in common with some of the more
dominant TS returned from the search procedure of step one. The $M$
shift points can be adaptively increased as new events with small
$d_{\E}^2$ are discovered. If the parameters used in the search of
step one are chosen wisely, then the initial list of shift points
should adequately cover the error region and the list of new error
events will be small.

It is common for the empirical variance, \eqref{eq: IS.IS var
estimate}, to underestimate the true variance in the high-SNR region
where the noise clouds are small.  This is because almost all of the
hits are intended hits, so the noise realizations don't venture
towards new error regions, where hits are likely to cause a large
$w(\yv)$. So, in the high-SNR region, when the initial TS list
excludes some dominant error events, these regions are being ignored
in $\hat{P_f}_{IS}$ and $\hat{V}$. For example, consider a standard
Monte Carlo simulation. At high SNR, we explore the $n$-dimensional
error region centered about the all-ones point (all-zeros codeword)
with $10^7$ noisy messages. Let the true $P_f$ be $10^{-8}$ at this
SNR. Thus with probability $(1-10^{-8})^{10^7}=0.9048$ we would not
get any hits in $10^7$ trials, resulting in an estimate of
$\hat{P_f}_{MC} = 0$ and an empirical variance also equal to zero.
Now, using \eqref{eq: f_star}, imagine placing the initial point of
reference at one of the $M$ shift points located between the
all-ones point and the dominant error event boundaries.  Since most
of the noise realizations fall much closer to the intended error
region, there will be many more hits in this error region and the
error contribution of at least those TS and codewords among the list
of $M$ shift points will be counted. Still, for all but relatively
short codes, this list will be incomplete and a large $\hat{V}$
could be obtained by having one or more noise realizations land in
error regions not included among the $M$ points in the initial list
that are closer to the all-ones vector than \textit{any} of our
shift points. This will produce a weight greater than one, which
will significantly increase $\hat{V}$. Although this will give a
large $\hat{V}$, it is still better to know that this previously
undiscovered error event exists. The alternative situation, where no
new error events are discovered, will produce a small $\hat{V}$, but
this is reminiscent of the Monte Carlo example above where the
regions of $\E$ not associated with our $M$ shift points are
ignored.

IS is no magical tool, and it really only helps when we know ahead
of time (steps one and two of the procedure) where the nearest error
regions are.  What we gain from using \eqref{eq: f_star} as our IS
$f^*$ is a significant reduction in the number of samples needed to
get a good estimate of $P_f$ compared to the method of finding the
$P_f$ contributed by each individual error event as detailed in
\cite{cavus-05, richardson}.  There is also a better chance of
accounting for the $P_f$ contributed by those error events which
were not explicitly enumerated with the search of step one in our
procedure, but are `close by' in $n$-dimensional decoding space to
some of the vectors that were in the list. Still, it must be
stressed that IS is a `dumb' procedure that helps when we already
have a very good list of dominant TS and codewords for the given
code.

\subsection{Complexity}  \label{sec: procedure_complexity}

It is difficult to attach a measure of complexity to our entire
procedure.  Traditionally, a `gain' metric is measured in an IS
simulation, usually a ratio of the number of samples required to
achieve a certain variance for $\hat{P_f}$ using Monte Carlo versus
applying IS.  This metric is essentially useless when applying IS to
the analysis of the error performance of large LDPC block codes. The
online variance estimator of \eqref{eq: IS.IS var estimate} which is
typically used to determine the number of samples required to
achieve a variance comparable to Monte Carlo for a given SNR is, as
outlined above, not reliable.  Another often overlooked aspect of
using IS to simulate decoding errors is the increase in the mean
number of iterations required to decode when $f^*$ causes most noise
realizations to fall near and in the error region, thus requiring
the decoder to `work harder' to find a valid codeword.  When the
list of mean-shift candidates in $f^*$ is large, the calculation of
the weight function is also time consuming.  So, ultimately, the
best choice of metric for measuring complexity will be the less
elegant - but more accurate - total compute-time required to run the
IS simulation. This will incorporate the total number of noise
samples, the extra iterations required for the MPA to decode
shifted-noise realizations, and the weight function calculation. To
measure complexity of the entire three-step process, also include
the time required to search for and determine relative dominance of
the initial list of TS from steps one and two.

\section{Simulation Results}\label{sec: results}

%Phase 1: $\epsilon_1 = 4, \gamma = 0.7, \frac{E_b}{N_o} = 7dB$,
%gives    TS, using all  $D_{\E}<60$, cuts list down to   TS.
% For (1008,504) no x2 num_trials = 39000

\begin{figure}[h]
\centerline{\hspace*{0.2in}\psfig{figure=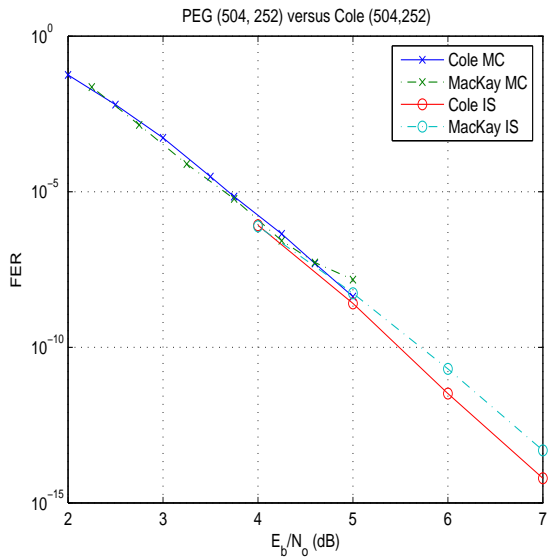,width=3.2in,height=3.0in}}
\caption{Comparison of (504,252) codes \label{fig: pegvscoleIS_504}}
\end{figure}

We now compare the error performance analysis results obtained from
the three-step procedure with a standard Monte Carlo estimate to see
if the results concur.  The Cole (504,252) code \cite{cole-06} with
girth eight and a progressive edge growth \cite{Hu_PEG-01} code on
the MacKay website \cite{Mackay_codes} will be used as a test case.
Figure \ref{fig: pegvscoleIS_504} has Monte Carlo data points up to
$E_b/N_o = 5$ dB for both codes.  The highest SNR point at 5 dB for
the Cole code involved $(3.15)10^9$ trials and 13 errors were
collected, so $\hat{P}_{f_{MC}}$ is not very accurate. It took about
6000 compute-hours on the cluster to obtain this point.  The IS data
for both codes, while slightly underestimating the true $P_f$, only
required 12 minutes to obtain the dominant TS list in step one,
negligible time to determine $d_{\E}^2$ for step two, and 2 hours
for the actual IS simulation.

The phenomenon of underestimating $P_f$ using the $f^*$ of step
three is the most troublesome weakness in our procedure.  Figure
\ref{fig: IS_1008_trials} shows three curves representing the IS
estimate of the (1008,504) Cole code. All three use the same $f^*$,
but the number of samples per SNR varies among 39000, 390000, and
$(3.9)10^6$. As more trials are performed, more of the error region
gets explored, thus increasing $\hat{P_f}_{IS}$.  This effect is
strongest at lower SNR where the noise clouds are larger and our
$f^*$ is less like the `optimal' $f^*$.

\begin{figure}[h]
\centerline{\hspace*{0.2in}\psfig{figure=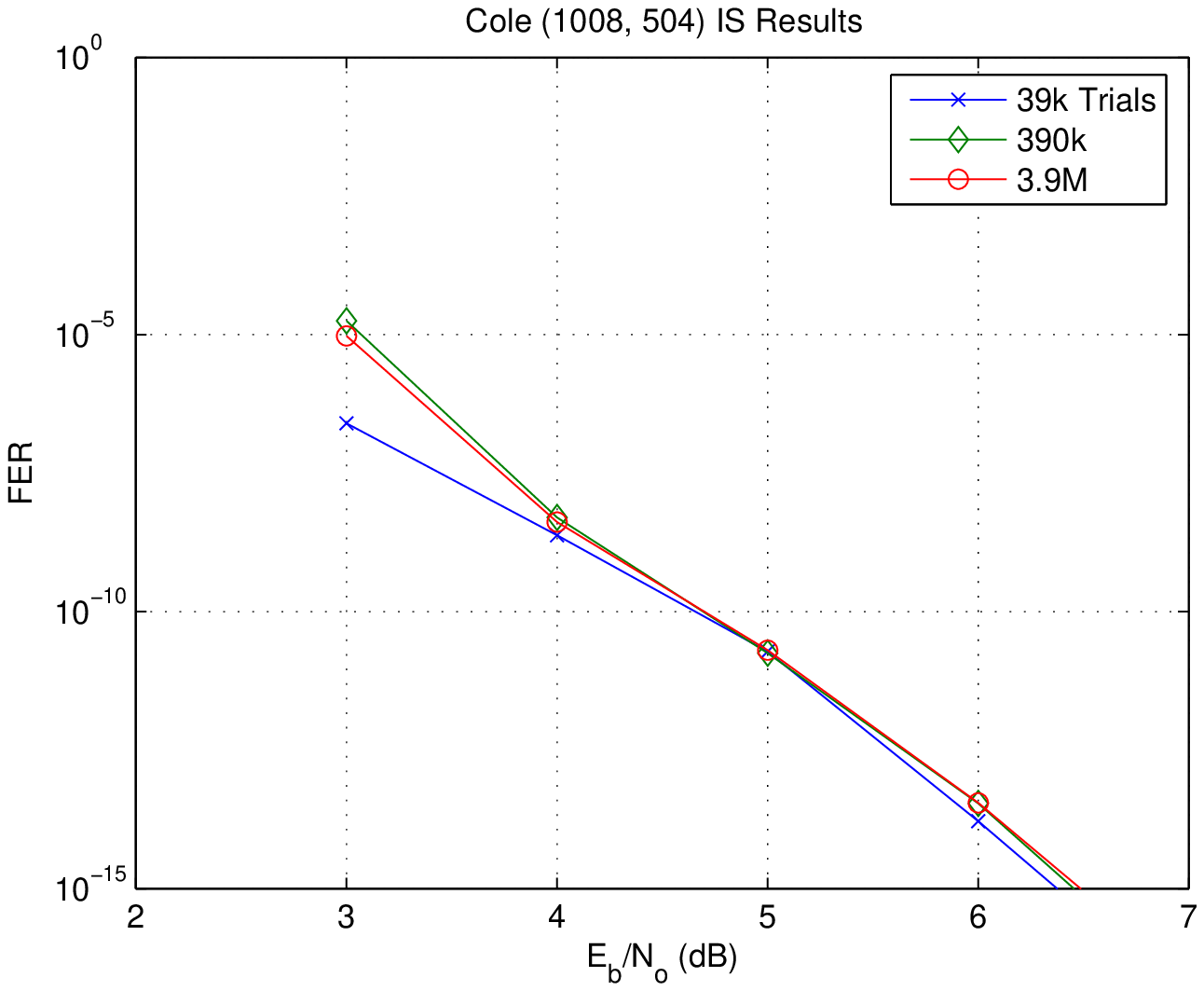,width=3.2in,height=3.0in}}
\caption{Number of trials effect on $\hat{P}_{f_{IS}}$ \label{fig:
IS_1008_trials}}
\end{figure}

In all of the following results, a maximum of 50 MPA iterations were
performed in decoding.  Table \ref{tab:phase_1_param} lists the
parameters used in the search phase of our procedure for a number of
different codes.  A value of $1-\gamma$ in the $\epsilon_2$ column
means $\epsilon_2$ was essentially not used and only the 4-bit
impulse with magnitude $\epsilon_1$ was used, and the rest of the
$n-4$ bits were scaled by $\gamma$. $\epsilon_2$ is usually only
required for larger codes.

\begin{table}
  \centering
\begin{tabular}{|c||l|l|l|l|r@{.}l|}
  \hline
%   &  &  &  &  &\\
     Code  & $\epsilon_1$ & $\epsilon_2$ & $\gamma$ & $E_b/N_o$ (dB) & \multicolumn{2}{|@{}c@{}|}{Time (Hrs)}\\
  \hline
(504,252) (Cole)& 3.6 & $1-\gamma$ & 0.8 & 5 & 0&2\\
(504,252) PEG (Hu)& 3.6 & $1-\gamma$ & 0.8 & 5 & 0&2\\
(603,301) Irregular (Dinoi)& 3.5 & $1-\gamma$ & 0.3 & 5 & 1&5 \\
(1008,504) (Cole) & 4 & $1-\gamma$ & 0.7 & 7 & 1&2 \\
(1008,504) PEG (Hu) & 4 & $1-\gamma$ & 0.7 & 7 & 1&2 \\
(2640,1320) (Margulis) & 5 & $1-\gamma$ & 0.3 & 6 & 8&2 \\
(4896,2448) (Ramanujan) & 5 & 2 & 0.4 & 6 & 24&0 \\
%(8192,4096) (Takeshita) & 6.25 & 4 & 0.45 & 6 & 320 \\
(1000,500) $\{4,8\}$ (MacKay) & 2.5 & $1-\gamma$ & 0.4 & 8 & 1&7 \\
  \hline
\end{tabular}
  \caption{Step one parameter values}\label{tab:phase_1_param}
\end{table}

The column labeled `Mean $d_{\E}^2$' in Table
\ref{tab:phase_2_param} is calculated over all $|EE|$ of the error
events, not just the ones that fall below the threshold, $d_{\E}^2 <
d_{\E_T}^2$. The `Time' column in Table \ref{tab:phase_3_param} is
on an AMD Athlon 2.2 GHz processor with 1 GByte RAM.

%\begin{figure*}
%\centerline{\subfigure[Case I]{
%\begin{figure}
\begin{table}
  \centering
\begin{tabular}{|c||l|l|l|l|l|}
  \hline
 %  &  &  &  &  &\\
     Code  & $|EE|$ & $d_{\E_T}^2$ & $|EE|_{< d_{\E_T}^2}$ & Min $d_{\E}^2$ & Avg $d_{\E}^2$  \\
  \hline
(504,252) (Cole)& 578 & - & 578 & 14.58 & 35.15 \\
(504,252) PEG (Hu)& 1954 & - & 1954 & 11.04 & 30.68 \\
(603,301) Irregular (Dinoi)& 10760 & 29 & 1499 & 15.47 & 49.29 \\
(1008,504) (Cole) & 750 & 60 & 390 & 21.45 & 57.20 \\
(1008,504) PEG (Hu) & 1700 & 60 & 1007 & 13.46 & 52.90 \\
(2640,1320) (Margulis) & 2640 & - & 2640 & 31.97 & 32.00 \\
(4896,2448) (Ramanujan) & 204 & - & 204 & 24.00 & 24.00 \\
(1000,500) $\{4,8\}$ (MacKay)  & 119 & 20 & 1 & 17.16 & 17.16 \\
%(8192,4096) (Takeshita) &  &  &  &  & \\
%$\{4,8\}$ (MacKay)  $D_{\E}$ = 17.1610
  \hline
\end{tabular}
  \caption{Step two parameter values}\label{tab:phase_2_param}
\end{table}
%\end{figure} } \hfil \subfigure[Margulis (2640,1320)]{
%\begin{figure}
\begin{table}
  \centering
\begin{tabular}{|c||l|l|r@{.}l|}
  \hline
  % &  &  &  \\
     Code  & \# Trials/SNR & $|EE|_{new}$ & \multicolumn{2}{|@{}c@{}|}{Time (Hrs)} \\
  \hline
(504,252) (Cole) & 195400 & 209 & 2&0 \\
(504,252) PEG & 115600 & 5649 & 2&0 \\
(603,301) Irregular (Dinoi) & 29980 & 3525 & 4&2 \\
(1008,504) (Cole) & 3900000 & 1574 & 78&8 \\
(1008,504) PEG (10,2)  & 302100 & 1842 & 5&0 \\
(2640,1320) (Margulis) & 208600 & 1001 & 19&5  \\
(4896,2448) (Ramanujan) & 204000 & 120 & 60&0 \\
(1000,500) $\{4,8\}$ (MacKay)  & 40000 & 2 & 2&0 \\
%(8192,4096) (Takeshita) &  &  & \\
  \hline
\end{tabular}
  \caption{Step three parameter values}\label{tab:phase_3_param}
\end{table}
%\end{figure} }} \hfil \caption{Simulation results} \label{fig_sim}
%\end{figure*}

\begin{figure*}
\centerline{\subfigure[Dinoi
(603,301)]{\includegraphics[width=3.2in]{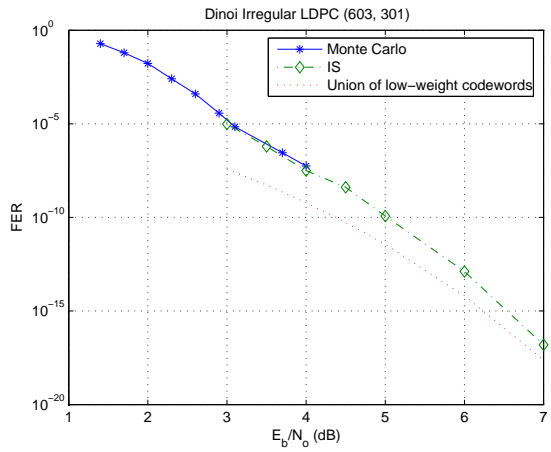}
\label{fig_first_case}} \hfil \subfigure[Margulis
(2640,1320)]{\includegraphics[width=3.2in]{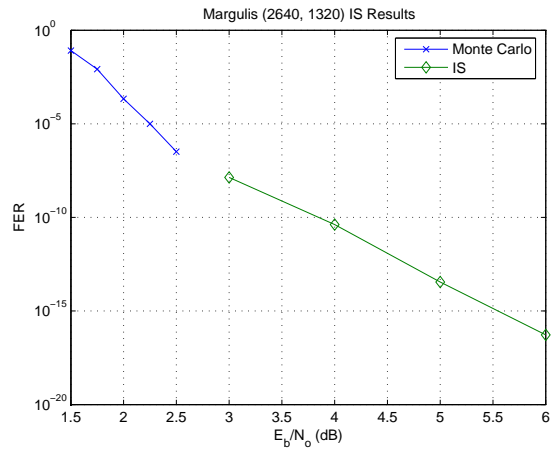}
\label{fig_second_case}}}\hfil {\subfigure[MacKay
(1000,500)]{\includegraphics[width=3.2in]{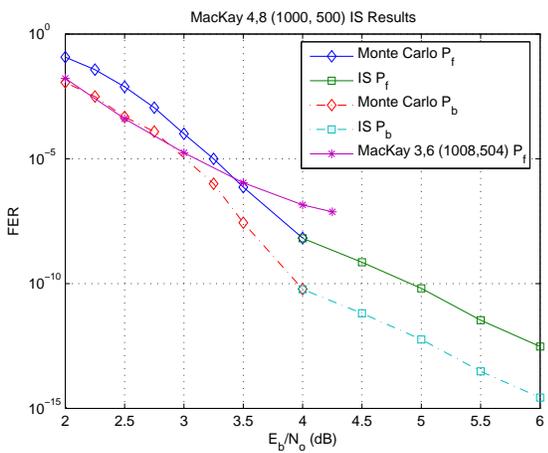}
\label{fig_third_case}} \hfil \subfigure[Ramanujan (4896,
2448)]{\includegraphics[width=3.2in]{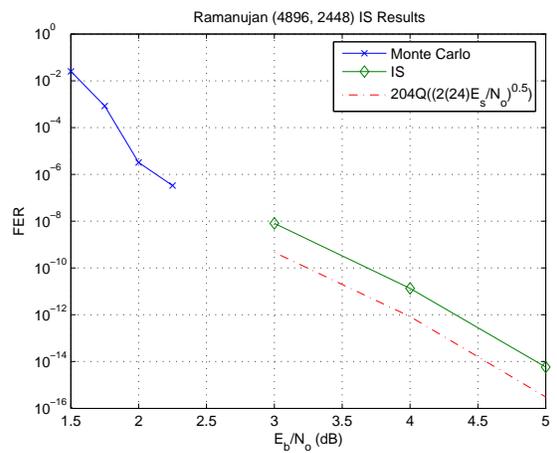}
\label{fig_fourth_case}}} \hfil \caption{Simulation results}
\label{fig_sim}
\end{figure*}

Applying the method to larger codes, an irregular code, and a
$\{4,8\}$ code highlights the generality of this three-step IS
method. The search step does not find all of the $(14,4)$ TS in the
Margulis (2640,1320) code \cite{margulis-82}, but these are all
supersets of $(12,4)$ TS which are all discovered.  When keeping
track of new error events in step three, many errors land in
$(14,4)$ TS, thus the $P_f$ associated with these $(14,4)$ TS is
accounted for in the total $P_f$. The IS results for the Margulis
code are shown in Figure \ref{fig_second_case}.  Step one found 204
codewords with a minimum $w_H$ of 24 in the Ramanujan (4896, 2448)
code \cite{Rosenthal-00}, and while this agrees with the number
found in \cite{Hu-04}, it's possible some were missed. The Ramanujan
code has an error floor dominated by valid codewords, but step three
of our procedure does find many non-codeword TS that are not
included among the initial list of 204 mean-shift points. As seen in
Figure \ref{fig_fourth_case} the total $P_f$ obtained from IS hovers
about an order of magnitude above the simple approximation
$204Q(\sqrt{2(24)E_s/N_o})$. Step three could be applied to the
quadratic permutation polynomial (8192,4096) code
\cite{takeshita-05} using the 3775 weight-52 codewords found in step
one as the shift points.  However, since $n$ is larger for this code
and the 3775 mean-shift points create a slow weight function
calculation, a quick alternative to a full IS simulation is to lower
bound $P_f$ by $3775Q(\sqrt{2(52)E_s/N_o})$. If step three were
performed, then a more accurate measurement of $P_f$ could be
obtained, i.e. $\hat{P_f}_{IS} \geq 3775Q(\sqrt{2(52)E_s/N_o})$.
These examples further highlight how step three of our procedure is
a better strategy for finding the $P_f$ of a code than the
previously proposed methods of considering only the $P_f$
contributed by each individual error event and then summing these.

Figure \ref{fig_third_case} shows how a MacKay $\{4,8\}$ (1000,500)
code has a major advantage over the traditionally considered
$\{3,6\}$ codes in the error floor region. The MacKay (1000,500)
code has some 4-cycles and its dominant error event is a single
(9,4) TS. No valid codewords were found. Even though the girth is
only four, the $\{4,8\}$ code has an error floor significantly lower
than the comparably-sized (1008,504) $\{3,6\}$ code with a girth of
six. So, although these extra cycles affect the decoder's threshold
region adversely, they do not degrade the high-SNR performance and
in fact the extra 33\% of edges in the graph improves the error
floor performance substantially.

The (603,301) irregular code developed in \cite{dinoi-05} has a low
error floor and the new method is an efficient way to measure this
code's performance.  The first step returns a list of 10760 TS, and
we use all 1499 of them which have $d_{\E}^2 < 29$.  Figure
\ref{fig_first_case} shows $\hat{P_f}_{IS}$ for this code.  The
curve matches well with the results in \cite{dinoi-05} up to where
the Monte Carlo data ends at $E_b/N_o = 4$ dB. The simulated curve
extending into the higher SNR region stays above the known lower
bound of $Q(\sqrt{2(15)E_s/N_o})$ caused by the code's single
weight-15 codeword.  These two checks reinforce our confidence in
the validity of the high SNR performance results given by the
three-step method described in this paper.

Finally, using (1008,504) $\{3,6\}$ codes, we also compare the
performance of the full belief propagation MPA with the min-sum MPA.
These surprising results are shown in Figure \ref{fig:
pegvscoleIS_1008}, where we see that at higher SNR (above 3.5 dB in
this case) the min-sum algorithm actually performs \textit{better}
than the much more complex belief propagation implementation!  This
behavior is evident for many codes that have an error floor
dominated by non-codeword TS.  The degree to which min-sum
outperforms BP is code dependent, but for some codes it is very
significant.  This surprising result is an example of the usefulness
of the low BER analysis tools presented in this paper.

%Error bars representing one standard deviation of our IS estimate,
%obtained by using the online variance estimator in equation
%\eqref{eq: IS.IS var estimate}, are plotted around the
%$\hat{P_f}_{IS}$ curve.

\begin{figure}[h]
\centerline{\hspace*{0.2in}\psfig{figure=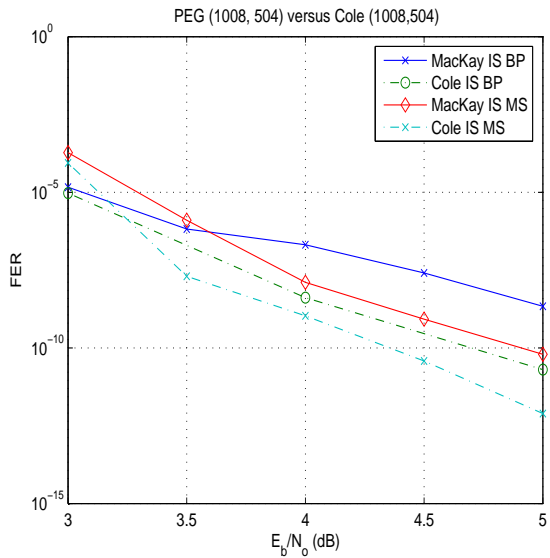,width=3.2in,height=3.0in}}
\caption{Comparison of (1008,504) codes \label{fig:
pegvscoleIS_1008}}
\end{figure}

%The IS results shown in this section should not be considered
%extremely accurate.  There were undoubtedly some fairly dominant TS
%left out of the weighted sum $f^*$, which would lead the IS estimate
%to be below the true value, and the few high-SNR Monte Carlo data
%points used for comparison purposes verify this. It is more accurate
%to describe the IS results as a tight lower bound on $P_f$.

\section{Conclusions}  \label{sec: conclusion}

The work presented here is very helpful in the analysis of LDPC code
error floors.  We developed a procedure that first uses a novel
search technique to find possibly dominant error events, then uses a
deterministic error impulse to determine which events in the initial
list are truly dominant error events, and then performs a
traditional mean-shifting IS technique to determine code performance
in the low bit error region.  This novel and general result has
applicability to the analysis of most classes of regular and
irregular LDPC codes and many decoders.

The procedure also provides the ability to accurately analyze and
iteratively adjust the code and decoder behavior in the error floor
region, a useful tool for applications that must have a guaranteed
very low bit error rate at a given SNR. One of the byproducts of
this research is the observation that the min-sum decoding algorithm
is just as good as, and in some cases better, than the full belief
propagation algorithm at sufficiently high SNR. In fact, this SNR is
usually just slightly higher than where Monte Carlo simulations
typically end, thus the result was always slightly out of reach of
previous researchers.  It was also shown that the class of regular
$\{4,8\}$ codes can provide a lower error floor than
comparable-length $\{3,6\}$ codes.

It is our belief that the methods outlined in this paper, while not
completely solving the problem of finding the low-weight TS spectrum
and error floor for the general class of long LDPC codes, will still
have a big impact on how researchers evaluate LDPC codes and
decoders in the high SNR region.  The work will provide a solid
foundation for others to build upon to attack even longer LDPC
codes.


\begin{thebibliography}{10}
\providecommand{\url}[1]{#1} \csname url@rmstyle\endcsname
\providecommand{\newblock}{\relax} \providecommand{\bibinfo}[2]{#2}
\providecommand\BIBentrySTDinterwordspacing{\spaceskip=0pt\relax}
\providecommand\BIBentryALTinterwordstretchfactor{4}
\providecommand\BIBentryALTinterwordspacing{\spaceskip=\fontdimen2\font
plus \BIBentryALTinterwordstretchfactor\fontdimen3\font minus
  \fontdimen4\font\relax}
\providecommand\BIBforeignlanguage[2]{{%
\expandafter\ifx\csname l@#1\endcsname\relax
\typeout{** WARNING: IEEEtran.bst: No hyphenation pattern has been}%
\typeout{** loaded for the language `#1'. Using the pattern for}%
\typeout{** the default language instead.}%
\else \language=\csname l@#1\endcsname \fi #2}}

\bibitem{Gallager}
R.G.Gallager, \emph{Low-Density Parity-Check Codes}.\hskip 1em plus
0.5em minus
  0.4em\relax MIT Press, 1963.

\bibitem{mackay-99}
D.~MacKay, ``Good error correcting codes based on very sparse
matrices,''
  \emph{IEEE Trans. Inf. Theory}, vol.~45, no.~3, pp. 399--431, Mar 1999.

\bibitem{cole-06}
C.~A. Cole, S.~G. Wilson, E.~K. Hall, and T.~R. Giallorenzi,
``Analysis and
  design of moderate length regular \textsc{LDPC} codes with low error
  floors,'' \emph{Conf. on Information Sciences and Systems}, Mar 2006.

\bibitem{richardson}
T.~Richardson, ``Error floors of \textsc{LDPC} codes,''
\emph{Allerton
  Conference}, 2001.

\bibitem{Srinivasan-02}
R.~Srinivasan, \emph{Importance Sampling - Applications in
Communications and
  Detection}.\hskip 1em plus 0.5em minus 0.4em\relax Springer-Verlag, 2002.

\bibitem{Smith-97}
P.~J. Smith, M.~Shafi, and H.~Gao, ``Quick simulation: A review of
importance
  sampling techniques in communications systems,'' \emph{IEEE JSAC}, vol.~15,
  no.~4, pp. 597--613, May 1997.

\bibitem{wu-95}
X.~Wu, ``\textsc{IS} - \textsc{B}lock and \textsc{C}onvolutional
  \textsc{C}odes,'' Master's thesis, University of Virginia, May 1995.

\bibitem{Xia-03}
B.~Xia and W.~Ryan, ``On importance sampling for linear block
codes,'' in
  \emph{Proc. IEEE ICC}, vol.~4, May 2003, pp. 2904--2908.

\bibitem{cavus-05}
E.~Cavus, C.~Haymes, and B.~Daneshrad, ``A highly efficient
importance sampling
  method for performance evaluation of \textsc{LDPC} codes at very low bit
  error rates,'' \emph{Submitted, IEEE Trans. Commun.}

\bibitem{Holzlohner-05}
R.~Holzlohner, A.~Mahadevan, C.~Menyuk, J.~Morris, and J.~Zweck,
``Evaluation
  of the very low \textsc{BER} of \textsc{FEC} codes using dual adaptive
  importance sampling,'' \emph{IEEE Comm. Letters}, vol.~2, Feb 2005.

\bibitem{tanner-81}
R.~M. Tanner, ``A recursive approach to low complexity codes,''
\emph{IEEE
  Trans. Inf. Theory}, vol.~27, no.~5, pp. 533--547, Sept 1981.

\bibitem{ryan}
W.~Ryan, ``\textsc{LDPC} tutorial,''
  \url{http://www.ece.arizona.edu/~ryan/New%20Folder/ryan-crc-ldpc-chap.pdf}.

\bibitem{Mackay_codes}
D.~Mackay, ``Mackay codes web site,''
  \url{http://www.inference.phy.cam.ac.uk/mackay/codes}.

\bibitem{Milenkovic-05}
O.~Milenkovic, E.~Soljanin, and P.~Whiting, ``Asymptotic spectra of
trapping
  sets in regular and irregular \textsc{LDPC} code ensembles,'' \emph{Submitted
  IEEE Trans. Inf. Theory}, 2005.

\bibitem{berrou-02}
C.~Berrou, S.~Vaton, M.~Jezequel, and C.~Douillard, ``Computing the
minimum
  distance of linear codes by the error impulse method,'' \emph{IEEE
  GlobeComm'02}, vol.~2, pp. 1017--1020, Nov 2002.

\bibitem{Hu-04}
X.-Y. Hu, M.~P.~C. Fossorier, and E.~Eleftheriou, ``On the
computation of the
  minimum distance of \textsc{LDPC} codes,'' \emph{IEEE ICC'04}, 2004.

\bibitem{takeshita-05}
O.~Y. Takeshita, ``A new construction for \textsc{LDPC} codes using
permutation
  polynomials over integer rings,'' \emph{Submitted, IEEE Trans. Inf. Theory},
  2005.

\bibitem{Rosenthal-00}
J.~Rosenthal and P.~Vontobel, ``Constructions of \textsc{LDPC} codes
using
  \textsc{R}amanujan graphs and ideas from \textsc{M}argulis,'' in \emph{Proc.
  38th Allerton Conf. Commun. (ICC)}.\hskip 1em plus 0.5em minus 0.4em\relax
  Monticello, Illinois, October 2000, pp. 248--257.

\bibitem{Song-06}
S.~Song, L.~Lan, S.~Lin, and K.~Abdel-Ghaffar, ``Construction of
quasi-cyclic
  \textsc{LDPC} codes based on the primitive elements of finite fields,''
  \emph{Conference on Info. Sciences and Systems}, Mar 2006.

\bibitem{shannon-59}
C.~E. Shannon, ``Probability of error for optimal codes in
\textsc{G}aussian
  channel,'' \emph{Bell Syst. Tech. J.}, vol.~38, pp. 611--656, 1959.

\bibitem{bucklewbook-04}
J.~A. Bucklew, \emph{Introduction to Rare Event Simulation}.\hskip
1em plus
  0.5em minus 0.4em\relax Springer, 2004.

\bibitem{Sadowsky-90}
J.~S. Sadowsky and J.~A. Bucklew, ``On large deviation theory and
  asymptotically efficient \textsc{M}onte-\textsc{C}arlo estimation,''
  \emph{IEEE Trans. Inf. Theory}, vol.~36, no.~3, pp. 579--588, May 1990.

\bibitem{Hu_PEG-01}
X.~Hu, E.~Eleftheriou, and D.~Arnold, ``Progressive edge-growth
\textsc{T}anner
  graphs,'' \emph{IEEE GlobeComm'01}, vol.~2, pp. 995--1001, Nov 2001.

\bibitem{margulis-82}
G.~A. Margulis, ``Explicit constructions of graphs without short
cycles and
  low-density codes,'' \emph{Combinatorica}, vol.~2, no.~1, pp. 71--78, 1982.

\bibitem{dinoi-05}
L.~Dinoi, F.~Sottile, and S.~Benedetto, ``Design of variable-rate
irregular
  \textsc{LDPC} codes with low error floors,'' in \emph{Proc. IEEE ICC},
  vol.~1, May 2005, pp. 647--651.

\end{thebibliography}
\end{document}